**Electrochemical impedance of electrodiffusion in charged medium under *dc* bias**

Juhyun Song[a], Edwin Khoo[a], and Martin Z. Bazant[a,b,*]

[a] Department of Chemical Engineering, Massachusetts Institute of Technology,
77 Massachusetts Avenue, Cambridge, Massachusetts 02139, United States

[b] Department of Mathematics, Massachusetts Institute of Technology,
77 Massachusetts Avenue, Cambridge, Cambridge, Massachusetts 02139, United States

**ABSTRACT**

An immobile charged species provides a charged medium for transport of charge carriers that is exploited in many applications, such as permselective membranes, doped semiconductors, biological ion channels, as well as porous media and microchannels with surface charges. In this paper, we theoretically study the electrochemical impedance of electrodiffusion in a charged medium by employing the Nernst-Planck equation and the electroneutrality condition with a background charge density. The impedance response is obtained under different *dc* bias conditions, extending above the diffusion-limiting bias. We find a transition in the impedance behavior around the diffusion-limiting bias, and present an analytical approximation for a weakly charged medium under an overlimiting bias.

* Corresponding author
    Postal address: 25 Ames Street, Room 66-458B, Cambridge, MA 02139, United States
    E-mail address: <u>bazant@mit.edu</u>
    Tel: +1-617-324-2036

**I. INTRODUCTION**

Transport of charged species relative to the surrounding medium is driven by gradients in their concentrations (i.e., diffusion) and in electric potential (i.e., migration or drift), which is therefore called electrodiffusion. The Nernst-Planck equation [1,2], also known as the drift-diffusion or the diffusion-migration equation, has successfully described electrodiffusion in various fields including batteries [3], fuel cells [4], semiconductors [5], solar cells [6], ion-exchange membranes [7-10], as well as biological systems [11-14], although its theoretical validity is restricted to dilute systems [15-17]. It is solved either under the electroneutrality condition in the thin-double-layer limit [18-20], or more generally under the electric potential that is determined self-consistently from the charge density via the Poisson equation governing electrostatics [17,21,22]. In the latter case, the model is called the Poisson-Nernst-Planck (PNP) set of equations, whose steady state solution leads to the Poisson-Boltzmann distribution and the Gouy-Chapman double layer model [23]. Recently, its transient response was thoroughly reviewed by Bazant and his colleagues [17]. Its impedance response has also been studied extensively by Macdonald and Franceschetti [20,22,24-27], Buck and Brumlev [28-30], Jamnik and Maier [31-33], Moya [34-36], and others [37-39] for various boundary conditions. When there is an excess amount of supporting electrolyte, or when there is a large disparity in mobilities of the charge carriers, the electrodiffusion models reduce to the neutral diffusion equation, or the Fick's law, exhibiting the Warburg behavior in its impedance response [20,40,41].

In many applications, there can be an immobile charged species that electrostatically interacts with the charge carriers. For example, in doped semiconductors, the dopants become ionized and provide immobile charges, either positive if the dopants are donors or negative if acceptors [5]. Permselective membranes, often used in desalination and chemical separations, typically have charged functional groups attached onto the polymer backbone chain [10,16,34,42,43]. In addition, ion channels in cell membranes are proteins that have a pore structure when open, whose inner surface consists of ionized groups [11,44,45]. The immobile charged species effectively provides a charged medium for the transport of charge carriers. Therefore, as suggested by Teorell, Meyer, and Sievers [8-10], the immobile charge species can be considered as a background charge density that is added to either the electroneutrality condition or the Poisson equation. Even for microchannels or porous media where the majority of charge carriers are out of the double layer on their local surface, it is recently shown that the surface charge can be treated as a homogenized background charge density in up-scaled macroscopic transport models [46].

Electric response of electrodiffusion in presence of an immobile charged species has been studied for the aforementioned applications in various set-ups including steady state [18,19,47-50], transient [50-55], and impedance [16,34,56-58]. When the immobile species has the opposite charge to the active charge carrier that carries the current at boundaries, the current may exceed the diffusion limit in steady state [18,19]. A depletion region is formed near the sink boundary where the concentration of charge carriers in the bulk electrolyte diminishes. The overlimiting current is sustained by either an extra conductivity provided by the charge carriers screening the immobile charges, or by electro-osmotic circulation and instability if the medium is fluid, in the depletion region [19,59]. The transient response under an overlimiting current is thoroughly studied by Zangle, Mani, and Santiago [52-55], Mani and Bazant [60] and Yaroshchuk [61], where they showed concentration polarization propagating as a shock wave. Also, Yan et al. [62] and Khoo and Bazant [50] have recently studied the linear sweep voltammetry of electrodiffusion in a charged medium with different boundary conditions. On the other hand, although there are a few studies on the impedance response in the context of a permselective membrane [56-58], a solar cell [6], and a microchannel [63,64], a general theory of the electrodiffusion impedance in a charged medium is not yet available to the best of our knowledge.

In this paper, we present a theoretical model for the electrochemical impedance of electrodiffusion in presence of an immobile charged species, and study its behavior under different *dc* biases. We consider a system that has two symmetric, oppositely charged charge carriers, which can be binary electrolytes in a liquid solution, electrons and holes in a semiconductor, or electrons and cations in a mixed conductor. No reaction or generation in the bulk is assumed, as well as negligible convection, as a first attempt to focus on the effect of the immobile charged species and the *dc* bias. Their contributions could be important in some applications, and the model in this paper should be modified accordingly [6,19,65,66]. We consider two configurations of boundary conditions: (*i*) a reservoir configuration in which one side of the system is exposed to a reservoir that maintains constant concentrations and potential and the other side is in contact with a selective boundary that only accepts the active charge carrier, and (*ii*) a symmetric configuration in which both sides are in contact with the selective boundaries. Our model isolates the response of the bulk electrodiffusion from the contributions of the interfaces and displacement current. For many applications where the electric double layer is much thinner than the system length scale (i.e., thin-double-layer limit) [17,67], the interfacial impedance appears well-separated in frequency from that of the bulk transport [14,43]. The displacement current also contributes at much higher frequencies. For a linearized response like impedance, their contributions could be obtained separately and added to the present model to examine the total cell impedance for the entire frequency range, as shown in our following study [68].

Beginning in Section II, we will first set up governing equations and boundary conditions, nondimensionalize variables, and then introduce a small perturbation to the system. In Section III, the zero-order terms are solved analytically to study the steady-state behavior. The first-order terms are then solved in Section IV, given the zero-order solutions, which provides the impedance response. We investigate the solutions for different combinations of immobile charge density and *dc* bias including those in the overlimiting regime. Asymptotic expressions for low- and high-frequency limits are obtained for the small- and large-bias regimes. Lastly, we also propose and validate a two-zone approximation for a large bias above the limiting current in Section V. The list of variables is given in Appendix A.

## II. MODEL

Consider a one-dimensional system from $x=0$ to $x=L$, containing two oppositely charged mobile species and another charged immobile species. It could be an unsupported liquid binary electrolyte in a permselective membrane with charged functional groups, or in a porous medium or a microchannel with surface charges. It could also be a solid-state system such as a semiconductor or a mixed ion-electron conductor containing ionized dopants. The Nernst-Planck equation is employed to describe electrodiffusion of the charge carriers. Assuming no generation and negligible convection,

$$\frac{\partial c_\pm}{\partial t} = -\frac{\partial F_\pm}{\partial x}, \tag{1}$$

$$F_\pm = -D\left(\frac{\partial c_\pm}{\partial x} \pm \frac{zec_\pm}{k_B T}\frac{\partial \phi}{\partial x}\right), \tag{2}$$

where $c_\pm(t,x)$ and $F_\pm(t,x)$ are the concentrations and the fluxes, respectively, of the positive and the negative charge carriers, and $\phi(t,x)$ is the electric potential. $t$ and $x$ are the time and the position variables. $e$, $k_B$, and $T$ are the electron charge, the Boltzmann constant, and the absolute temperature, respectively. Here we assumed a symmetric charge number, $z$, and a symmetric chemical diffusivity, $D$. Although $D$ is generally a function of concentrations, we consider it constant for simplicity. For a porous medium or a microchannel, $D$ is the effective diffusivity corrected by the porosity and the tortuosity [69], and $\phi$ is the electric potential in the solution phase.

Assuming thin double layers, we isolate the quasi-neutral bulk and adopt the local electroneutrality condition to replace the Poisson equation. The immobile charged species appears as a background charge in the neutrality condition,

$$0 = zec_+ - zec_- + \rho, \tag{3}$$

where $\rho$ is the charge density of the immobile species. For a porous medium or microchannel, Equation (3) means macroscopic electroneutrality, which is valid as long as the pores are larger than a couple of nanometers [70,71]. Then, $\rho$ is the macroscopically homogenized density of the surface charge; $\rho = a_v \sigma_s / \epsilon_p$, where $a_v$ is the volumetric area, $\sigma_s$ is the surface charge, and $\epsilon_p$ is the porosity (unity for a microchannel) [46,50,72]. This condition constrains $c_+$ and $c_-$ to vary exactly in phase with each other, keeping a constant offset $\rho/ze$ in bulk. The results in this paper are illustrated by assigning the positive charge carriers to be the active carrier. This choice is arbitrary and the results are symmetric in $\rho$, if the negative charge carriers become the active carriers.

Let us define the dimensionless variables: $\tilde{x} = x/L$, $\tilde{t} = Dt/L^2$, $\tilde{\rho} = \rho/2zec_0$, $\tilde{\phi} = ze\phi/k_BT$, and $\tilde{c} = c_-/c_0 = c_+/c_0 + 2\tilde{\rho}$, where $c_0$ is the initial anion concentration at equilibrium. Then by Equations (2) and (3), the dimensionless fluxes ($\tilde{F}_\pm = LF_\pm/Dc_0$) become

$$\tilde{F}_+ = -\frac{\partial \tilde{c}}{\partial \tilde{x}} - (\tilde{c} - 2\tilde{\rho})\frac{\partial \tilde{\phi}}{\partial \tilde{x}}, \tag{4}$$

$$\tilde{F}_- = -\frac{\partial \tilde{c}}{\partial \tilde{x}} + \tilde{c}\frac{\partial \tilde{\phi}}{\partial \tilde{x}}. \tag{5}$$

Upon plugging the flux equations into the conservation equations given by Equation (1), we add and subtract the conservation equations of the positive and the negative charge carriers. Thus, the following dimensionless governing equations are obtained for $\tilde{c}(\tilde{t},\tilde{x})$ and $\tilde{\phi}(\tilde{t},\tilde{x})$.

$$\frac{\partial \tilde{c}}{\partial \tilde{t}} = \frac{\partial^2 \tilde{c}}{\partial \tilde{x}^2} - \tilde{\rho}\frac{\partial^2 \tilde{\phi}}{\partial \tilde{x}^2}, \tag{6}$$

$$0 = \frac{\partial}{\partial \tilde{x}}\left((\tilde{c} - \tilde{\rho})\frac{\partial \tilde{\phi}}{\partial \tilde{x}}\right). \tag{7}$$

The first equation describes conservation of the total number of charge carriers, whereas the latter equation describes conservation of the net electric charge. By assuming the local neutrality condition in Equation (3), we are able to reduce the number of concentration variables, and drop the time derivative for the conservation of net electric charge.

As for the boundary conditions, we consider two common cell configurations as shown in FIG 1. In both configurations, a potential ($-\tilde{V} = -zeV/k_BT$) is applied at the selective boundary at $\tilde{x}=1$, where only the active carrier (the positive charge carrier in this paper) can pass the current and the other carrier is completely blocked ($\tilde{F}_- = 0$). Since we are isolating the bulk electrodiffusion, the boundary conditions are defined inside the transport medium, not involving charge transfer kinetics or double layer charging dynamics at the interfaces. Influence of boundary impedance is discussed in detail elsewhere by Macdonald using the Chang-Jaffe kinetics [73] and by ourselves using the Butler-Volmer kinetics [68]. Therefore, in this work, the boundary conditions at $\tilde{x}=1$ are

$$\tilde{\phi} = -\tilde{V}, \tag{8}$$

$$\frac{d\tilde{c}}{d\tilde{x}} - \tilde{c}\frac{d\tilde{\phi}}{d\tilde{x}} = 0. \tag{9}$$

In the reservoir configuration, FIG 1 (a), an ideal reservoir maintains $\tilde{c}$ and $\tilde{\phi}$ constant on the other side ($\tilde{x}=0$). We set $\tilde{\phi}$ at $\tilde{x}=0$ to be zero as a reference value. Therefore, the boundary conditions for the reservoir configuration at $\tilde{x}=0$ are

$$\tilde{c} = 1, \tag{10}$$

$$\tilde{\phi} = 0. \tag{11}$$

In the symmetric configuration, FIG 1 (b), the boundary at $\tilde{x}=0$ has the same semi-blocking condition ($\tilde{F}_- = 0$) as well, identical to the boundary at $\tilde{x}=1$. The boundary conditions for the symmetric configuration at $\tilde{x}=0$ are

$$\frac{d\tilde{c}}{d\tilde{x}} - \tilde{c}\frac{d\tilde{\phi}}{d\tilde{x}} = 0, \tag{12}$$

$$\tilde{\phi} = 0. \tag{13}$$

The two blocking conditions for the negative charge carrier in the symmetric configuration do not specify a unique solution. To obtain a unique solution, the total number of the negative charge carriers should be conserved, which adds the following integral constraint for the symmetric configuration.

$$\int_0^1 \tilde{c}\,d\tilde{x} = 1. \tag{14}$$

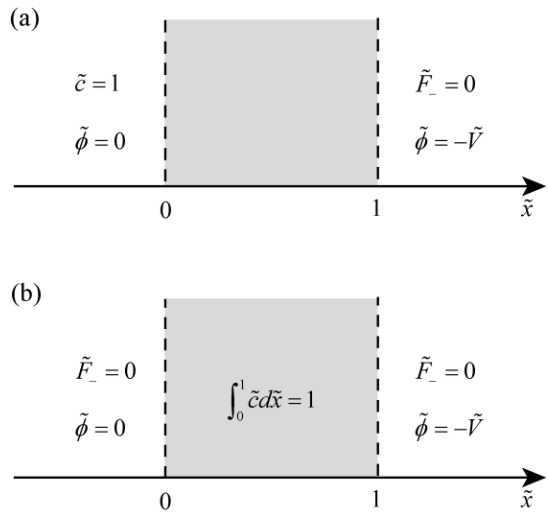

FIG 1. Two cell configurations and their boundary conditions: (a) reservoir and (b) symmetric configurations.

Regardless of configuration, the dimensionless system is governed by two dimensionless parameters, $\tilde{\rho}$ and $\tilde{V}$, which represent the charge density of the immobile species and the magnitude of the applied bias respectively. It is also possible to specify the applied bias by the current ($j$) as well. Disregarding the displacement current, the current comes from the charge carrier fluxes: $j = ze(F_+ - F_-)$. When scaled by the diffusion limiting current ($j_{\text{lim}} = 2zeDc_0/L$), the dimensionless current ($\tilde{j} = j/j_{\text{lim}}$) can be obtained by

$$\tilde{j} = -(\tilde{c} - \tilde{\rho})\frac{\partial \tilde{\phi}}{\partial \tilde{x}}. \tag{15}$$

$\tilde{j}$ is constant throughout the system domain due to the local neutrality condition in Equation (3). In this paper, we refer to $\tilde{j}$ to specify the magnitude of the *dc* bias, rather than $\tilde{V}$, because it gives a more relevant scale to electrodiffusion.

The system is perturbed by a small sinusoidal stimulus with a frequency $\tilde{\omega} = \omega L^2/D$, in either $\tilde{V}$ or $\tilde{j}$ to calculate the impedance. When the amplitude is small enough, variables can then be expanded according to the perturbation theory.

$$\begin{aligned}
\tilde{c}(\tilde{x},\tilde{t}) &= \tilde{c}^{(0)}(\tilde{x}) + \epsilon \tilde{c}^{(1)}(\tilde{x}) e^{i\tilde{\omega}\tilde{t}} + O(\epsilon^2), \\
\tilde{\phi}(\tilde{x},\tilde{t}) &= \tilde{\phi}^{(0)}(\tilde{x}) + \epsilon \tilde{\phi}^{(1)}(\tilde{x}) e^{i\tilde{\omega}\tilde{t}} + O(\epsilon^2), \\
\tilde{j}(\tilde{t}) &= \tilde{j}^{(0)} + \epsilon \tilde{j}^{(1)} e^{i\tilde{\omega}\tilde{t}} + O(\epsilon^2), \\
\tilde{V}(\tilde{t}) &= \tilde{V}^{(0)} + \epsilon \tilde{V}^{(1)} e^{i\tilde{\omega}\tilde{t}} + O(\epsilon^2),
\end{aligned} \tag{16}$$

where $\epsilon$ is an arbitrary small number. Other variables may be expanded in the same manner as well. Upon substitution of the expanded variables into Equations (6) − (15), the collection of $O(1)$ terms can be firstly solved for the reference steady state under a *dc* bias. Given the steady state solution, the collection of $O(\epsilon)$ terms can be solved for the perturbation around the steady state, which is then used for calculating the impedance.

## III. STEADY STATE

Impedance is measured by applying a perturbation around a reference steady state. By achieving different steady states by *dc* bias, we can study nonlinear behavior of charge carriers via impedance spectroscopy [74]. To interpret such results, we need to incorporate the steady state in an impedance model. Although the steady state solution under *dc* bias has already been studied in Ref. [19,50], we revisit its behavior for completeness of the paper. The steady state solution is obtained by solving the $O(1)$ terms, and steps involved in obtaining an analytical solution are presented in Appendix B. For the reservoir configuration, the solution is

$$\tilde{j}^{(0)} = 1 - \left(e^{-\tilde{V}^{(0)}} + \tilde{\rho}\tilde{V}^{(0)}\right), \tag{17}$$

$$\tilde{c}^{(0)} = \begin{cases} -\tilde{\rho}\, W_0\left[-\tilde{\rho}^{-1}\exp\left\{-\tilde{\rho}^{-1}\left(1-\tilde{j}^{(0)}\tilde{x}\right)\right\}\right] & ,\text{if } \tilde{\rho}<0, \\ 1-\tilde{j}^{(0)}\tilde{x} & ,\text{if } \tilde{\rho}=0, \\ -\tilde{\rho}\, W_{-1}\left[-\tilde{\rho}^{-1}\exp\left\{-\tilde{\rho}^{-1}\left(1-\tilde{j}^{(0)}\tilde{x}\right)\right\}\right] & ,\text{if } \tilde{\rho}>0, \end{cases} \tag{18}$$

$$\tilde{\phi}^{(0)} = \log\left(\tilde{c}^{(0)}\right), \tag{19}$$

where $W_0$ and $W_{-1}$ are the principal (upper) and the lower branches of the Lambert W function, respectively [75].

FIG 2 (a) shows $\tilde{j}^{(0)}$ against $\tilde{V}^{(0)}$ for $\tilde{\rho} \in \{-0.1, -0.01, 0, 0.01, 0.1\}$. Without the immobile charge ($\tilde{\rho}=0$), $\tilde{j}^{(0)}$ saturates at the diffusion limiting current ($\tilde{j}^{(0)}=1$). If the immobile species and the active carriers are co-charged ($\tilde{\rho}>0$), $\tilde{j}^{(0)}$ is limited by $1-\tilde{\rho}\left(2-\log(2\tilde{\rho})\right)$, at which $\tilde{c}^{(0)}$ reaches $2\tilde{\rho}$ at $\tilde{x}=1$. When charge transfer kinetics is considered, the current limit cannot be reached with a finite potential bias since the charge transfer resistance diverges due to the diminishing concentration at the boundary. On the other hand, if the immobile species and the active carrier are counter-charged ($\tilde{\rho}<0$), $\tilde{j}^{(0)}$ can exceed the limiting current above which it keeps a constant slope of $-\tilde{\rho}$. In the overlimiting regime, the current is sustained by the active carriers that screen the immobile charge.

In FIG 2 (b) and (c), $\tilde{c}^{(0)}$ and $\tilde{\phi}^{(0)}$ are plotted under a varying *dc* bias for $\tilde{\rho}=-0.01$. As a bias is applied, a concentration gradient develops, and it becomes steeper as $\tilde{j}^{(0)}$ increases. Below the limiting current, $\tilde{c}^{(0)}$ drops almost linearly along $\tilde{x}$, while $\tilde{\phi}^{(0)}$ shows a steeper drop closer to the boundary at $\tilde{x}=1$. On the other hand, above the limiting current, the depletion region appears near the boundary at $\tilde{x}=1$, where $\tilde{c}^{(0)}$ diminishes and $\tilde{\phi}^{(0)}$ shows a steep linear decrease. Therefore, the contribution of electric migration should be larger compared to that of diffusion in the depletion region. Also, notice that the depletion region grows as $\tilde{j}^{(0)}$ increases. Our MATLAB script is available on GitHub ([https://github.com/JuhyunSong/Impedance_2019.git](https://github.com/JuhyunSong/Impedance_2019.git)) which plots the steady state solution for the reservoir configuration with a given combination of $\tilde{\rho}$ and $\tilde{j}^{(0)}$. When $\tilde{\rho}$ becomes more negative, the transition in $\tilde{c}^{(0)}$ and $\tilde{\phi}^{(0)}$ becomes smoother which makes the depletion region less distinctive. An analytical approximation for the overlimiting regime is discussed in Section V.

For the symmetric configuration, an analytical solution can be also obtained. Steps involved in solving the $O(1)$ terms are presented in Appendix B as well. The solution for the symmetric configuration is

$$\tilde{j}^{(0)} = \alpha\left(1 - e^{-\tilde{V}^{(0)}}\right) - \tilde{\rho}\tilde{V}^{(0)}, \tag{20}$$

$$\tilde{c}^{(0)} = \begin{cases} -\tilde{\rho}\,W_0\left[-\alpha\tilde{\rho}^{-1}\exp\left\{-\tilde{\rho}^{-1}\left(\alpha - \tilde{j}^{(0)}\tilde{x}\right)\right\}\right] & , \text{if } \tilde{\rho} < 0, \\ \alpha - \tilde{j}^{(0)}\tilde{x} & , \text{if } \tilde{\rho} = 0, \\ -\tilde{\rho}\,W_{-1}\left[-\alpha\tilde{\rho}^{-1}\exp\left\{-\tilde{\rho}^{-1}\left(\alpha - \tilde{j}^{(0)}\tilde{x}\right)\right\}\right] & , \text{if } \tilde{\rho} > 0, \end{cases} \tag{21}$$

$$\tilde{\phi}^{(0)} = \log\left(\frac{\tilde{c}^{(0)}}{\alpha}\right), \tag{22}$$

where $\alpha$ is an intermediate parameter such that

$$\alpha = \frac{(1+\tilde{\rho})\left(1 - e^{-\tilde{V}^{(0)}}\right) + \sqrt{(1+\tilde{\rho})^2\left(1 - e^{-\tilde{V}^{(0)}}\right)^2 - 2\tilde{\rho}\tilde{V}^{(0)}\left(1 - e^{-2\tilde{V}^{(0)}}\right)}}{1 - e^{-2\tilde{V}^{(0)}}}. \tag{23}$$

It can be inferred by Equation (22) that $\alpha = \tilde{c}^{(0)}(\tilde{x}=0)$.

FIG 2 (d) shows $\tilde{j}^{(0)}$ against $\tilde{V}^{(0)}$ for $\tilde{\rho} \in \{-0.1, -0.01, 0, 0.01, 0.1\}$. While its general behavior is similar to FIG 2 (a), the diffusion limiting current in the symmetric configuration is twice of the reservoir configuration. The current is limited by 2 if $\tilde{\rho}=0$, or by $\alpha - \tilde{\rho}(2 - \log(2\tilde{\rho}/\alpha))$ if $\tilde{\rho} > 0$, where $\alpha$ is implicitly determined by putting $-\tilde{V}^{(0)} = \log(2\tilde{\rho}/\alpha)$ in Equation (23). On the other hand, the overlimiting regime is accessible if $\tilde{\rho} < 0$. In FIG 2 (e) and (f), $\tilde{c}^{(0)}$ and $\tilde{\phi}^{(0)}$ are plotted under a varying dc bias for $\tilde{\rho}=-0.01$. When $\tilde{j}^{(0)} > 0$, $\tilde{c}^{(0)}$ begins to have a gradient and $\tilde{c}^{(0)}$ at $\tilde{x}=1$ drops, which results in $\tilde{c}^{(0)}$ building up near the boundary at $\tilde{x}=0$ due to the integral constraint in Equation (14). It results in a steeper gradient, which enhances the transport and makes the diffusion limiting current twice higher than that of the reservoir configuration. Other than that, $\tilde{c}^{(0)}$ and $\tilde{\phi}^{(0)}$ show similar behaviors to those of the reservoir configuration (FIG 2 (b) and (c)). Our MATLAB script is available on GitHub (https://github.com/JuhyunSong/Impedance_2019.git) which plots the steady state solution for the symmetric configuration with a given combination of $\tilde{\rho}$ and $\tilde{j}^{(0)}$. An analytical approximation for the overlimiting regime in the symmetric configuration is discussed in Section V as well. The steady state solutions are used in calculating the $O(\epsilon)$ perturbation solutions in the next Section.

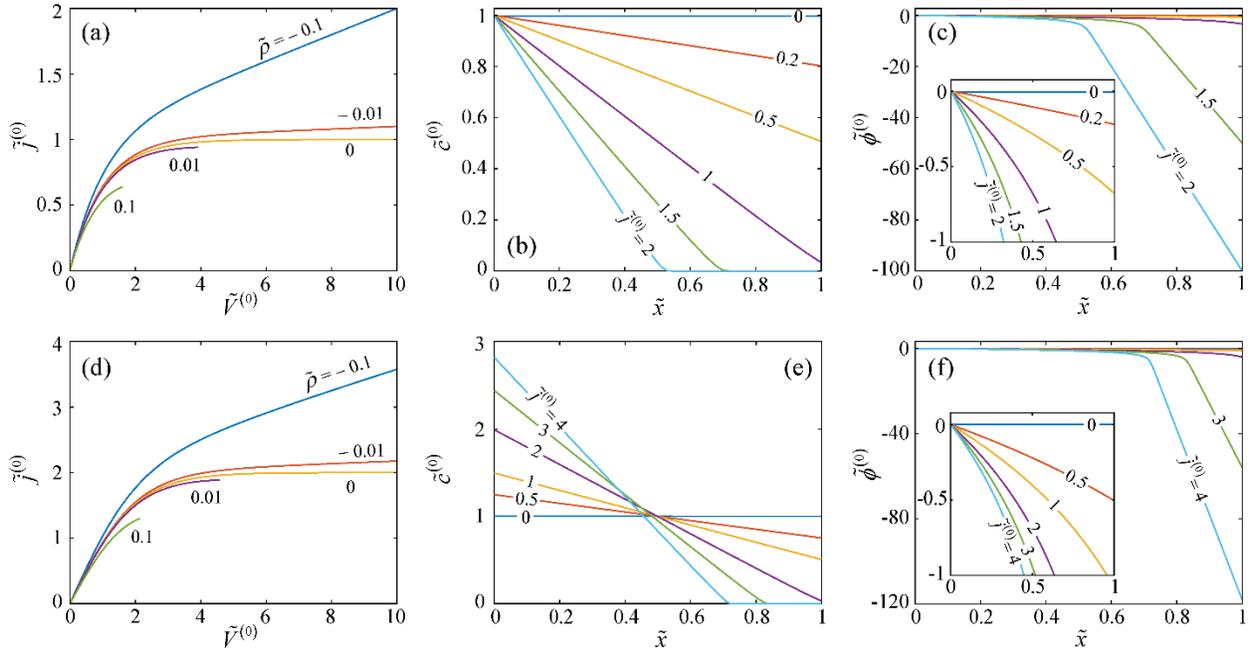

FIG 2. Steady state solution for the reservoir configuration: (a) Current-voltage curves for different $\tilde{\rho}$. (b) Concentration and (c) potential distributions for $\tilde{\rho}=-0.01$ under different *dc* biases. For the symmetric configuration: (d) Current-voltage curves for different $\tilde{\rho}$. (e) Concentration and (f) potential distributions for $\tilde{\rho}=-0.01$ under different *dc* biases.

## IV. IMPEDANCE

While the $O(1)$ solution corresponds to the reference steady state under a *dc* bias, the $O(\epsilon)$ terms describe the perturbation around it, which leads to the impedance solution. The $O(\epsilon)$ governing equations are

$$i\tilde{\omega}\tilde{c}^{(1)} = \frac{d^2\tilde{c}^{(1)}}{d\tilde{x}^2} - \tilde{\rho}\frac{d^2\tilde{\phi}^{(1)}}{d\tilde{x}^2}, \tag{24}$$

$$0 = \left(\tilde{c}^{(0)} - \tilde{\rho}\right)\frac{d^2\tilde{\phi}^{(1)}}{d\tilde{x}^2} + \frac{d\tilde{c}^{(0)}}{d\tilde{x}}\frac{d\tilde{\phi}^{(1)}}{d\tilde{x}} + \frac{d\tilde{\phi}^{(0)}}{d\tilde{x}}\frac{d\tilde{c}^{(1)}}{d\tilde{x}} + \frac{d^2\tilde{\phi}^{(0)}}{d\tilde{x}^2}\tilde{c}^{(1)}. \tag{25}$$

For the reservoir configuration, the $O(\epsilon)$ boundary conditions are

$$\tilde{c}^{(1)} = 0 \quad \text{at } \tilde{x} = 0, \tag{26}$$

$$\tilde{\phi}^{(1)} = 0 \quad \text{at } \tilde{x} = 0, \tag{27}$$

$$\tilde{\phi}^{(1)} = -\tilde{V}^{(1)} \quad \text{at } \tilde{x} = 1, \tag{28}$$

$$\frac{d\tilde{c}^{(1)}}{d\tilde{x}} = \tilde{c}^{(1)} \frac{d\tilde{\phi}^{(0)}}{d\tilde{x}} + \tilde{c}^{(0)} \frac{d\tilde{\phi}^{(1)}}{d\tilde{x}} \quad \text{at } \tilde{x} = 1. \tag{29}$$

This boundary value problem involves the $O(1)$ solution derived in the previous Section. Since they are not elementary functions, Equations (24) – (29) should be solved numerically. We post our MATLAB script on GitHub (https://github.com/JuhyunSong/Impedance_2019.git), which solves the boundary value problem and calculates the impedance for any combinations of $\tilde{j}^{(0)}$ and $\tilde{\rho}$. When the $O(\epsilon)$ solution is obtained, combined with the $O(1)$ solution, the impedance ($\tilde{Z} = 2Dc_0 Z/k_B TL$) is calculated by

$$\tilde{Z} = \frac{\tilde{V}^{(1)}}{\tilde{j}^{(1)}} = \frac{\tilde{V}^{(1)}}{-\left(\tilde{c}^{(0)} - \tilde{\rho}\right)\left(d\tilde{\phi}^{(1)}/dx\right) - \tilde{c}^{(1)}\left(d\tilde{\phi}^{(0)}/dx\right)}. \tag{30}$$

FIG 3 presents $\tilde{Z}$ under different biases on the complex plane, for the reservoir configuration with $\tilde{\rho} = -0.01$. Below the diffusion-limiting bias ($\tilde{j}^{(0)} < 1$) as shown in FIG 3 (a), the impedance of bulk electrodiffusion appears as a semicircle that grows in both low-frequency and high-frequency limits with increasing bias. At the high-frequency limit, it is dominated by the $O(\epsilon)$ conduction with the steady state conductivity that leads to the pure resistive behavior. The capacitive contribution comes from the fluctuation in $O(\epsilon)$ conductivity under the steady state potential gradient, which turns back to the resistive behavior at the lower-frequency limit when the conductivity fluctuates in phase with $\tilde{V}^{(1)}$ and $\tilde{j}^{(1)}$. Around the diffusion-limiting bias ($\tilde{j}^{(0)} = 1$), the semicircle becomes suppressed, and then the impedance appears more like a finite-length Warburg element under a bias above the diffusion limit ($\tilde{j}^{(0)} > 1$), as shown in FIG 3 (b). Its overall magnitude also starts converging above the limit. Such response comes from the conduction in the depletion region and the diffusion out of the depletion region, under an overlimiting bias. Detailed discussion on this regime is presented in the next Section with an analytical approximation. Without any bias, the impedance shrinks to a pure resistance of $(1-\tilde{\rho})^{-1}$.

The transition along increasing bias becomes more apparent by plotting the limiting behaviors as functions of the applied bias $\tilde{j}^{(0)}$. FIG 4 (a) shows the low-frequency resistance, $\tilde{R}_L = \tilde{Z}(\tilde{\omega} \to 0)$, and the high-frequency resistance, $\tilde{R}_H = \tilde{Z}(\tilde{\omega} \to \infty)$. Both increase exponentially in the underlimiting regime ($\tilde{j}^{(0)} < 1$), and then converge to $\tilde{\rho}^{-1}$ in the overlimiting regime ($\tilde{j}^{(0)} > 1$). Asymptotic expressions of $\tilde{R}_L$ and $\tilde{R}_H$ are derived in Appendix C for the small-bias limit ($\tilde{j}^{(0)} \ll 1$) and in Appendix D for the large-bias limit ($\tilde{j}^{(0)} \gg 1$).

$$\tilde{R}_L\left(\tilde{j}^{(0)} \ll 1\right) \simeq \frac{1-\tilde{\rho}}{(1-\tilde{\rho})^2 - \tilde{j}^{(0)}}, \tag{31}$$

$$\tilde{R}_L\left(\tilde{j}^{(0)} \gg 1\right) \simeq \frac{1}{-\tilde{\rho}}, \tag{32}$$

$$\tilde{R}_H\left(\tilde{j}^{(0)} \ll 1\right) \simeq -\frac{(1-\tilde{\rho})}{\tilde{j}^{(0)}}\log\left(\frac{(1-\tilde{\rho})^2 - \tilde{j}^{(0)}}{(1-\tilde{\rho})^2}\right), \tag{33}$$

$$\tilde{R}_H\left(\tilde{j}^{(0)} \gg 1\right) \simeq \left(\frac{1}{-\tilde{\rho}}\right)\frac{\tilde{j}^{(0)}-1}{\tilde{j}^{(0)}}. \tag{34}$$

They are compared to the numerical limiting behaviors in FIG A1. In addition, FIG 4 (b) shows the local phase angle at the high-frequency limit $\theta_H$. In the underlimiting regime, $\theta_H$ starts by −90° which corresponds to the semicircle shape on the complex plane, FIG 3 (a). It then approaches to −45° in the overlimiting regime as the impedance transitions to the finite-length Warburg shape. If $\tilde{\rho} \geq 0$, $\tilde{R}_L$ and $\tilde{R}_H$ diverges before the diffusion-limiting bias. The maximum current bias in the reservoir configuration is 1 if $\tilde{\rho}=0$, or $1-\tilde{\rho}(2-\log(2\tilde{\rho}))$ if $\tilde{\rho}>0$, as discussed in Section III.

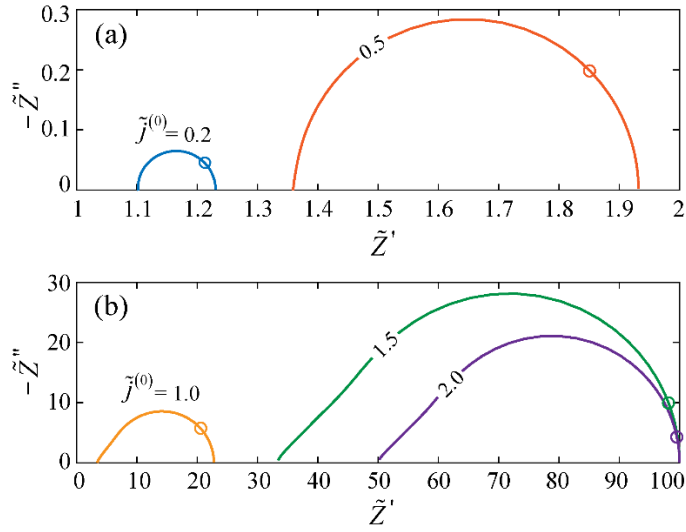

FIG 3. $\tilde{Z}$ in the reservoir configuration with $\tilde{\rho}=-0.01$ under different *dc* biases: (a) $\tilde{j}^{(0)}=0.2$ and 0.5. (b) $\tilde{j}^{(0)}=1.0$, 1.5, and 2.0. The circular makers are $\tilde{Z}(\tilde{\omega}=1)$.

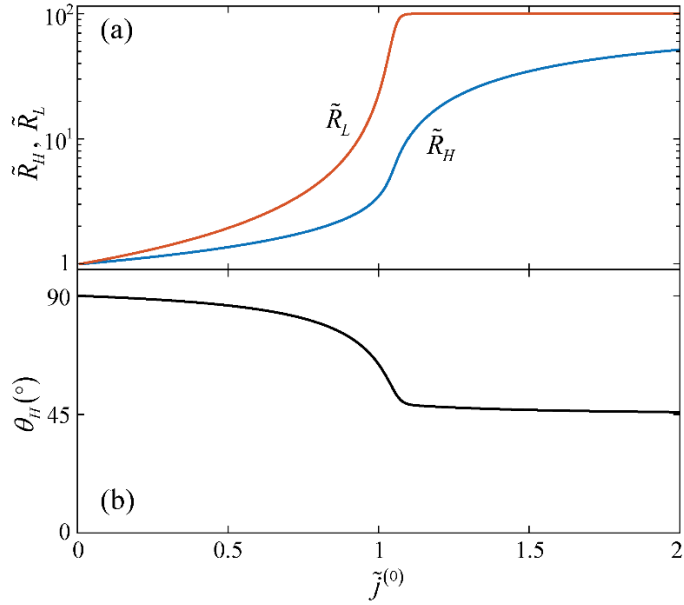

FIG 4. Limiting behaviors of $\tilde{Z}$ as functions of $\tilde{j}^{(0)}$ in the reservoir configuration with $\tilde{\rho} = -0.01$: (a) $\tilde{R}_L$ and $\tilde{R}_H$, and (b) $\theta_H$.

In a similar approach, the impedance for the symmetric configuration can be obtained. Its $O(\epsilon)$ governing equations are the same with the reservoir case, Equations (24) and (25). The $O(\epsilon)$ boundary conditions for the symmetric configuration are

$$\tilde{\phi}^{(1)} = 0 \quad \text{at } \tilde{x} = 0, \tag{35}$$

$$\frac{d\tilde{c}^{(1)}}{d\tilde{x}} = \tilde{c}^{(1)} \frac{d\tilde{\phi}^{(0)}}{d\tilde{x}} + \tilde{c}^{(0)} \frac{d\tilde{\phi}^{(1)}}{d\tilde{x}} \quad \text{at } \tilde{x} = 0, \tag{36}$$

$$\tilde{\phi}^{(1)} = -\tilde{V}^{(1)} \quad \text{at } \tilde{x} = 1, \tag{37}$$

$$\frac{d\tilde{c}^{(1)}}{d\tilde{x}} = \tilde{c}^{(1)} \frac{d\tilde{\phi}^{(0)}}{d\tilde{x}} + \tilde{c}^{(0)} \frac{d\tilde{\phi}^{(1)}}{d\tilde{x}} \quad \text{at } \tilde{x} = 1, \tag{38}$$

$$\int_0^1 \tilde{c}^{(1)} d\tilde{x} = 0. \tag{39}$$

To solve Equations (24), (25), and (35)−(39) with a typical solver for boundary value problems, another field variable is introduced, $\tilde{y}^{(1)}(\tilde{x}) = \int_0^{\tilde{x}} \tilde{c}^{(1)}(s) ds$, which adds another governing equation, $d\tilde{y}^{(1)}/d\tilde{x} = \tilde{c}^{(1)}$. Then the integral boundary condition, Equation (39), is replaced by two new boundary conditions for $\tilde{y}^{(1)}(\tilde{x})$: $\tilde{y}^{(1)}(0) = 0$ and $\tilde{y}^{(1)}(1) = 1$. The implementation can be found in our MATLAB script posted on GitHub (https://github.com/JuhyunSong/Impedance_2019.git), which solves the $O(\epsilon)$ problem and calculates the impedance for the symmetric configuration with any combinations of $\tilde{j}^{(0)}$ and $\tilde{\rho}$.

FIG 5 shows $\tilde{Z}$ under different biases for the symmetric configuration with $\tilde{\rho}=-0.01$, and FIG 6 shows the transition of its limiting behaviors along increasing $\tilde{j}^{(0)}$. Like in the reservoir configuration, $\tilde{Z}$ appears similar to a semicircle on the complex plane below the diffusion-limit bias ($\tilde{j}^{(0)} < 2$), whose $\tilde{R}_L$ and $\tilde{R}_H$ increase exponentially with increasing $\tilde{j}^{(0)}$ as shown in FIG 6 (a). However, the semicircle is tilted by a small angle at the high-frequency limit due to the integral constraint in the symmetric configuration, Equation (14). Therefore, $\theta_H$ starts less negative than -90° in FIG 6 (b). Then $\tilde{Z}$ shows a transition around the diffusion-limiting bias ($\tilde{j}^{(0)} = 2$), and it appears like a finite-length Warburg element in the overlimiting regime ($\tilde{j}^{(0)} > 2$) as shown in FIG 5 (b). Similar to the transition observed in the reservoir configuration, $\tilde{R}_L$ and $\tilde{R}_H$ converge to $\tilde{\rho}^{-1}$, and $\theta_H$ converges to -45° in the overlimiting regime. Asymptotic expressions of $\tilde{R}_L$ and $\tilde{R}_H$ for the symmetric configuration are derived in Appendix C for the small-bias limit ($\tilde{j}^{(0)} \ll 2$) and in Appendix D for the large-bias limit ($\tilde{j}^{(0)} \gg 2$).

$$\tilde{R}_L\left(\tilde{j}^{(0)} \ll 2\right) \simeq \frac{1-\tilde{\rho}}{2(1-\tilde{\rho})^2 - \tilde{j}^{(0)}} + \frac{1-\tilde{\rho}}{2(1-\tilde{\rho})^2 + \tilde{j}^{(0)}}, \tag{40}$$

$$\tilde{R}_L\left(\tilde{j}^{(0)} \gg 2\right) \simeq \frac{1}{-\tilde{\rho}}\left(1 - \frac{1}{\sqrt{2\tilde{j}^{(0)}}}\right), \tag{41}$$

$$\tilde{R}_H\left(\tilde{j}^{(0)} \ll 2\right) \simeq -\frac{(1-\tilde{\rho})}{\tilde{j}^{(0)}} \log\left(\frac{-\tilde{j}^{(0)} + 2(1-\tilde{\rho})^2}{\tilde{j}^{(0)} + 2(1-\tilde{\rho})^2}\right), \tag{42}$$

$$\tilde{R}_H\left(\tilde{j}^{(0)} \gg 2\right) \simeq \frac{1}{-\tilde{\rho}} \frac{\tilde{j}^{(0)} - \sqrt{2\tilde{j}^{(0)}}}{\tilde{j}^{(0)}}. \tag{43}$$

They are compared to the numerical limiting behaviors in FIG A1. The current bias $\tilde{j}^{(0)}$ is limited by 2 if $\tilde{\rho}=0$, or by $\alpha - \tilde{\rho}(2-\log(2\tilde{\rho}/\alpha))$ if $\tilde{\rho}>0$, as discussed in Section III, and $\tilde{R}_L$ and $\tilde{R}_H$ diverge before the limiting current bias. Regardless of $\tilde{\rho}$, $\tilde{Z}$ becomes a pure resistance of $(1-\tilde{\rho})^{-1}$ without any bias, same as the reservoir configuration.

Notice our model focuses on the bulk electrodiffusion without considering the contributions of the interfaces and displacement current. Each of these additional contributions would yield another relaxation behavior at higher frequencies usually well-separated from the bulk features [14,20]. Also, if the diffusivities of positive and negative charge carriers are different, another finite-length Warburg appears around the bulk diffusion frequency starting from the unbiased condition. While we have isolated its contribution out in this paper by assuming identical diffusivities, the extra Warburg coming from unequal diffusivities could appear merged with the bulk impedance studied in this paper [20,37,39]. In our following study, we apply the model and the solution methods to a more general set up with unequal diffusivities and charge numbers [68].

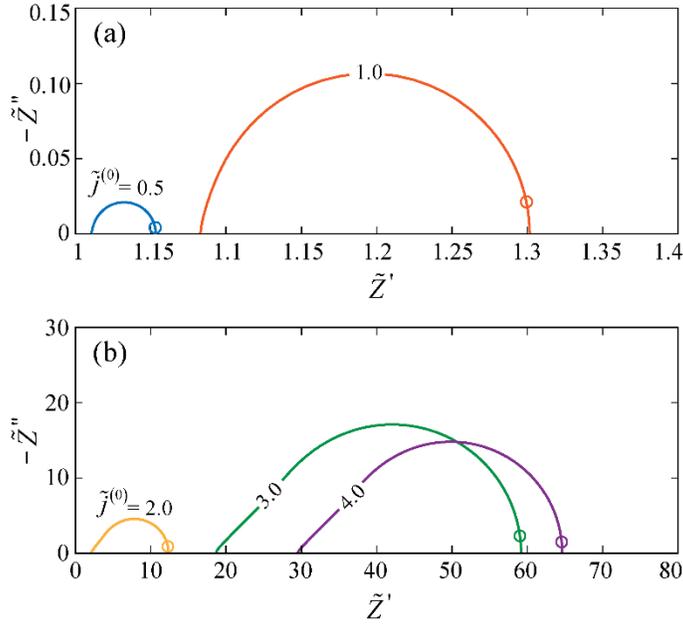

FIG 5. $\tilde{Z}$ in the symmetric configuration with $\tilde{\rho}=-0.01$ under different *dc* biases: (a) $\tilde{j}^{(0)}=0.5$ and 0.1. (b) $\tilde{j}^{(0)}=2.0$, 3.0, and 4.0. The circular makers are $\tilde{Z}(\tilde{\omega}=1)$.

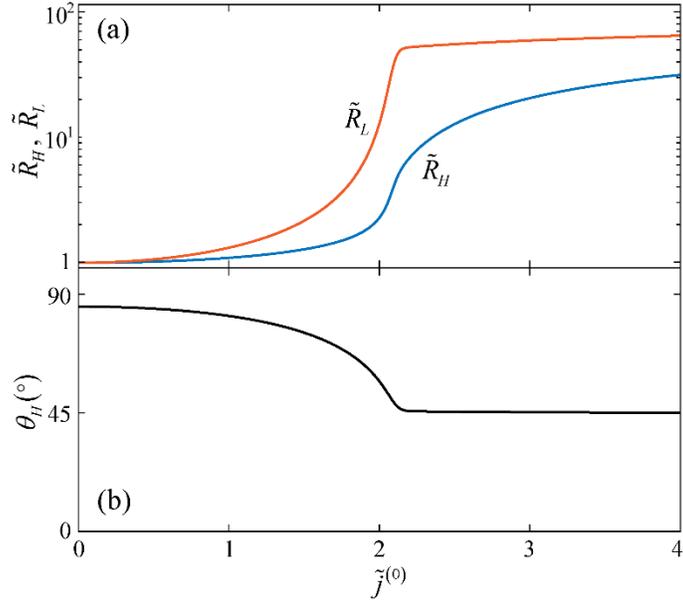

FIG 6. Limiting behaviors of $\tilde{Z}$ as functions of $\tilde{j}^{(0)}$ in the symmetric configuration with $\tilde{\rho}=-0.01$: (a) $\tilde{R}_L$ and $\tilde{R}_H$, and (b) $\theta_H$.

## V. TWO-ZONE APPROXIMATION

In this section, we present an analytical approximation for a counter-charged system with a low density of the immobile charged species and with a large bias above the diffusion-limiting bias. It is motivated by recognizing that $\tilde{c}^{(0)}(\tilde{x})$ and $\tilde{\phi}^{(0)}(\tilde{x})$ show two distinct zones under such conditions as shown in FIG 7 (a) and (b). The two zones become more apparent by defining the diffusion and the conduction currents, $\tilde{j}_D^{(0)}$ and $\tilde{j}_C^{(0)}$, for steady state:

$$\tilde{j}_D^{(0)} = -\frac{d\tilde{c}^{(0)}}{d\tilde{x}}, \qquad (44)$$

$$\tilde{j}_C^{(0)} = \tilde{\rho}\frac{d\tilde{\phi}^{(0)}}{d\tilde{x}}, \qquad (45)$$

which add up to the total current; $\tilde{j}_D^{(0)} + \tilde{j}_C^{(0)} = \tilde{j}^{(0)}$. FIG 7 (c) shows how contributions of $\tilde{j}_D^{(0)}$ and $\tilde{j}_C^{(0)}$ change along $\tilde{x}$. From $\tilde{x} = 0$, expanding to a position around the middle in the Figure, $\tilde{c}^{(0)}$ drops linearly while $\tilde{\phi}^{(0)}$ changes little compared to the rest of the space. In this zone, $\tilde{j}_D^{(0)}$ dominates, and we define it as the *diffusion zone* with length $\tilde{l}_D$ that depends on $\tilde{j}^{(0)}$ and $\tilde{\rho}$. On the other hand, from $\tilde{x} = \tilde{l}_D$ to 1, $\tilde{c}^{(0)}$ stays near zero while $\tilde{\phi}^{(0)}$ shows a large linear drop. Since $\tilde{j}_C^{(0)}$ dominates here, we define this zone as the *conduction zone* with length $\tilde{l}_C$ ($\tilde{l}_D + \tilde{l}_C = 1$), which is referred to as the depletion region in a lot of literature on the overlimiting current. In this approximation, the Nernst-Planck model is reduced to a pure diffusion equation in the diffusion zone, and to a pure conduction equation in the conduction zone. We will derive the analytical solutions for steady state as well as the impedance in both of the reservoir and symmetric configurations. Its validity is then discussed over a range of $\tilde{j}^{(0)}$ and $\tilde{\rho}$ in the last part.

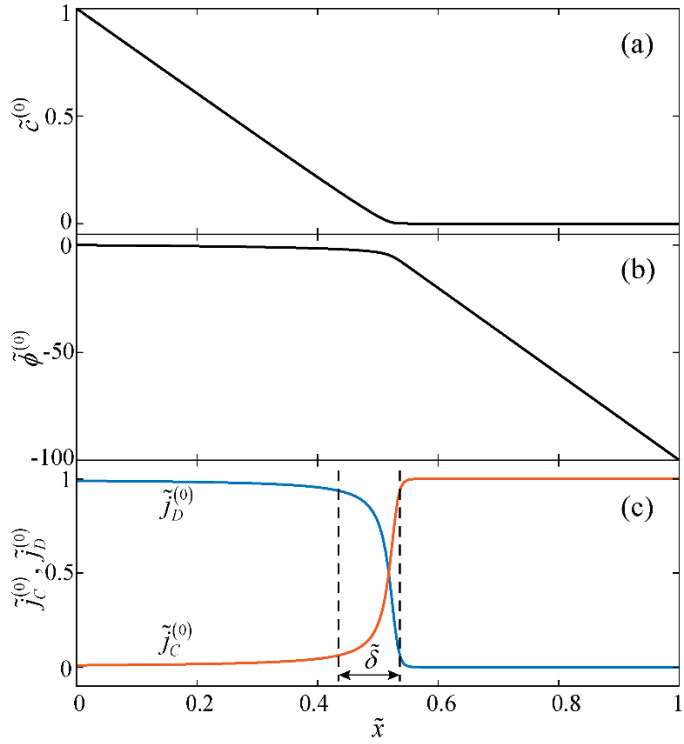

FIG 7. Spatial distributions of (a) $\tilde{c}^{(0)}$, (b) $\tilde{\phi}^{(0)}$, and (c) $\tilde{j}_D^{(0)}$ and $\tilde{j}_C^{(0)}$ at an overlimiting current ($\tilde{j}^{(0)} = 2.0$) in the reservoir configuration with $\tilde{\rho} = -0.01$.

For the reservoir configuration in steady state, the approximated concentration $\tilde{c}_{TZ}^{(0)}$ is obtained by solving $d^2\tilde{c}_{TZ}^{(0)}/d\tilde{x}^2 = 0$ in the diffusion zone with boundary conditions $\tilde{c}_{TZ}^{(0)} = 1$ at $\tilde{x} = 0$ and $\tilde{c}_{TZ}^{(0)} = 0$ at $\tilde{x} = \tilde{l}_D$. On the other hand, $\tilde{c}_{TZ}^{(0)}$ is kept zero throughout the conduction zone. Therefore,

$$\tilde{c}_{TZ}^{(0)} = \begin{cases} 1 - \dfrac{\tilde{x}}{\tilde{l}_D} &, \text{if } 0 \leq \tilde{x} \leq \tilde{l}_D, \\ 0 &, \text{if } \tilde{l}_D < \tilde{x} \leq 1, \end{cases} \tag{46}$$

where $\tilde{l}_D$ is determined such that the concentration gradient drives $\tilde{j}^{(0)}$ in the diffusion zone by Equation (44): $\tilde{l}_D = 1/\tilde{j}^{(0)}$. The approximated potential $\tilde{\phi}_{TZ}^{(0)}$ is then obtained by solving the pure conduction equation, $\tilde{\rho} d\tilde{\phi}_{TZ}^{(0)}/d\tilde{x} = \tilde{j}^{(0)}$, in the conduction zone. Since the potential drop in the diffusion zone is significantly smaller than that in the conduction zone, we set $\tilde{\phi}_{TZ}^{(0)} = 0$ in the diffusion zone. Therefore,

$$\tilde{\phi}_{TZ}^{(0)} = \begin{cases} 0 & \text{, if } 0 \leq \tilde{x} \leq \tilde{l}_D, \\ \dfrac{\tilde{j}^{(0)}}{\tilde{\rho}}(\tilde{x} - \tilde{l}_D) & \text{, if } \tilde{l}_D < \tilde{x} \leq 1. \end{cases} \qquad (47)$$

$\tilde{c}_{TZ}^{(0)}$ and $\tilde{\phi}_{TZ}^{(0)}$ are compared to the exact solutions in FIG 8 (a) and (b), respectively. When $|\tilde{\rho}|$ is small and $\tilde{j}^{(0)}$ is large above the diffusion-limit, the two-zone approximation shows good agreement with the exact steady state solutions.

Impedance approximation can then be obtained by separately calculating impedance of the two zones. For impedance of the diffusion zone, we first calculate $\tilde{c}_{TZ}^{(1)}$ at $\tilde{x} = \tilde{l}_D$ given $\tilde{j}^{(1)}$. $\tilde{c}_{TZ}^{(1)}$ is obtained by solving $d^2 \tilde{c}_{TZ}^{(1)}/d\tilde{x}^2 = i\tilde{\omega}\tilde{c}_{TZ}^{(1)}$ in the diffusion zone with boundary conditions $\tilde{c}_{TZ}^{(1)} = 0$ at $\tilde{x} = 0$ and $d\tilde{c}_{TZ}^{(1)}/d\tilde{x} = -\tilde{j}^{(1)}$ at $\tilde{x} = \tilde{l}_D$.

$$\tilde{c}_{TZ}^{(1)} = -\tilde{j}^{(1)} \tilde{l}_D \frac{\sinh(\sqrt{i\tilde{\omega}}\tilde{x})}{\sqrt{i\tilde{\omega}}\tilde{l}_D \cosh(\sqrt{i\tilde{\omega}}\tilde{l}_D)}, \qquad (48)$$

for $0 \leq \tilde{x} \leq \tilde{l}_D$ (i.e., in the diffusion zone). To calculate $\tilde{\phi}_{TZ}^{(1)}(\tilde{x} = \tilde{l}_D)$, we consider the $O(\epsilon)$ terms of Equations (15) and (44): $(\tilde{c}_{TZ}^{(0)} - \tilde{\rho}) d\tilde{\phi}_{TZ}^{(1)}/d\tilde{x} = d\tilde{c}_{TZ}^{(1)}/d\tilde{x}$. Since $\tilde{c}_{TZ}^{(0)}$ approaches zero at $\tilde{x} = \tilde{l}_D$, we can integrate $d\tilde{\phi}_{TZ}^{(1)}/d\tilde{c}_{TZ}^{(1)} = -\tilde{\rho}^{-1}$ from the unperturbed condition (i.e., $\tilde{\phi}_{TZ}^{(1)} = 0$ and $\tilde{c}_{TZ}^{(1)} = 0$), and obtain

$$\frac{\tilde{\phi}_{TZ}^{(1)}(\tilde{x} = \tilde{l}_D)}{\tilde{c}_{TZ}^{(1)}(\tilde{x} = \tilde{l}_D)} = \frac{1}{-\tilde{\rho}}. \qquad (49)$$

Then, impedance in the diffusion zone ($\tilde{Z}_{DZ}$) can be obtained by

$$\tilde{Z}_{DZ} = \frac{-\tilde{\phi}_{TZ}^{(1)}(\tilde{x} = \tilde{l}_D)}{\tilde{j}^{(1)}} = \frac{-\tilde{\phi}_{TZ}^{(1)}(\tilde{x} = \tilde{l}_D)}{\tilde{c}_{TZ}^{(1)}(\tilde{x} = \tilde{l}_D)} \frac{\tilde{c}_{TZ}^{(1)}(\tilde{x} = \tilde{l}_D)}{\tilde{j}^{(1)}} = \frac{\tilde{l}_D}{-\tilde{\rho}} \frac{\tanh(\sqrt{i\tilde{\omega}}\tilde{l}_D)}{\sqrt{i\tilde{\omega}}\tilde{l}_D}, \qquad (50)$$

which turns out to be a finite-length Warburg element scaled by $\tilde{l}_D/(-\tilde{\rho})$. Also, its characteristic frequency corresponds to diffusion with length $\tilde{l}_D$. On the other hand, impedance in the conduction zone ($\tilde{Z}_{CZ}$) is a pure resistor with a conductivity of $-\tilde{\rho}$ and a length of $\tilde{l}_C$. Therefore, the overall impedance in the two-zone approximation is

$$\tilde{Z}_{TZ} = \tilde{Z}_{CZ} + \tilde{Z}_{DZ} = \frac{\tilde{l}_C}{-\tilde{\rho}} + \frac{\tilde{l}_D}{-\tilde{\rho}} \frac{\tanh(\sqrt{i\tilde{\omega}}\tilde{l}_D)}{\sqrt{i\tilde{\omega}}\tilde{l}_D}, \qquad (51)$$

where $\tilde{l}_C = \left(\tilde{j}^{(0)} - 1\right)\big/\tilde{j}^{(0)}$ and $\tilde{l}_D = 1/\tilde{j}^{(0)}$. In FIG 8 (c), $\tilde{Z}_{TZ}$ is compared to the exact $\tilde{Z}$ obtained by a numerical solution obtained in Section IV for the reservoir configuration. The two-zone approximation provides an accurate approximation in impedance for a weakly charged system ($\tilde{\rho} = -0.01$) under an overlimiting current bias. It also infers that the finite-length Warburg feature corresponds to the diffusion zone and the resistor corresponds to the conduction zone. As $\tilde{j}^{(0)}$ increases above the diffusion-limit, the conduction zone pushes out the diffusion zone, and the finite-length Warburg part becomes smaller while the resistor becomes larger, keeping the overall low-frequency limit the same.

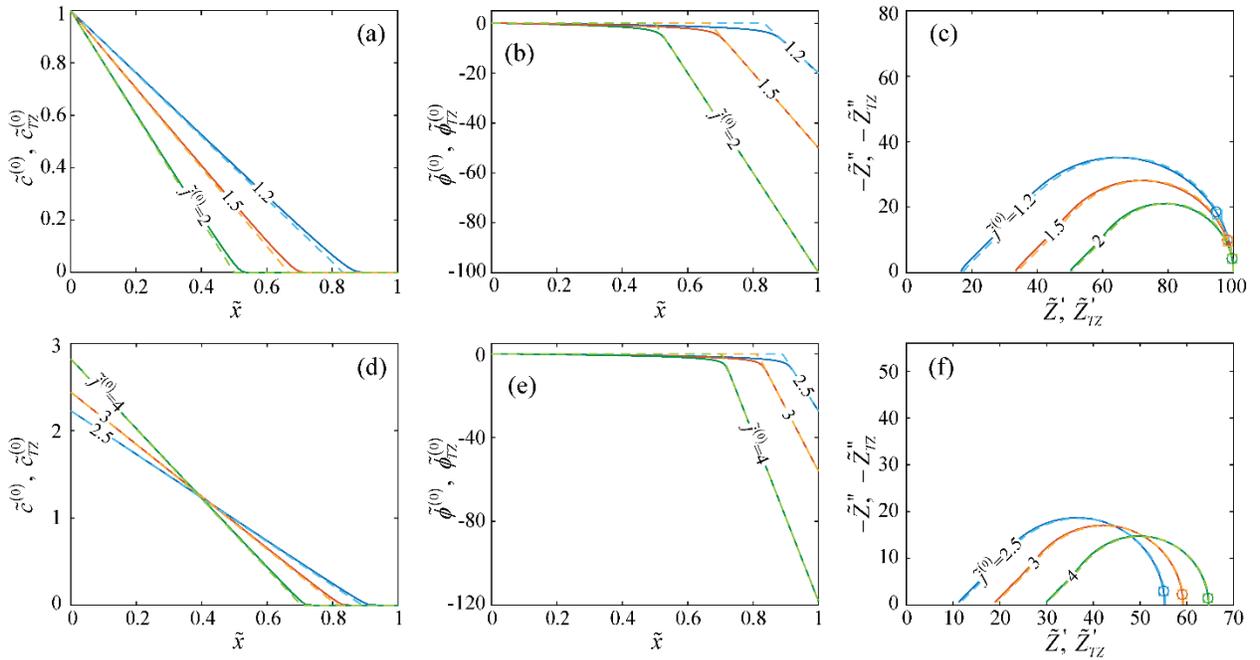

FIG 8. Comparison of the two-zone approximation with the numerical solution. For the reservoir configuration with $\tilde{\rho} = -0.01$: (a) $\tilde{c}^{(0)}$ and $\tilde{c}_{TZ}^{(0)}$, (b) $\tilde{\phi}^{(0)}$ and $\tilde{\phi}_{TZ}^{(0)}$ along $\tilde{x}$, as well as (c) $\tilde{Z}$ and $\tilde{Z}_{TZ}$, under different overlimiting biases. For the symmetric configuration with $\tilde{\rho} = -0.01$: (a) $\tilde{c}^{(0)}$ and $\tilde{c}_{TZ}^{(0)}$, (b) $\tilde{\phi}^{(0)}$ and $\tilde{\phi}_{TZ}^{(0)}$ along $\tilde{x}$, as well as (c) $\tilde{Z}$ and $\tilde{Z}_{TZ}$, under different overlimiting biases. The numerical solutions are in solid lines, and the approximations are in dashed lines. In (c) and (f), the circular and the square markers are $\tilde{Z}(\tilde{\omega} = 1)$ and $\tilde{Z}_{TZ}(\tilde{\omega} = 1)$, respectively.

For the symmetric configuration, the approximation can be obtained in a similar approach. For the steady state, $\tilde{c}_{TZ}^{(0)}$ is obtained by solving $d^2\tilde{c}_{TZ}^{(0)}/d\tilde{x}^2 = 0$ in the diffusion zone with boundary conditions $\tilde{c}_{TZ}^{(0)} = \alpha$ at $\tilde{x} = 0$ and $\tilde{c}_{TZ}^{(0)} = 0$ at $\tilde{x} = \tilde{l}_D$. On the other hand, $\tilde{c}_{TZ}^{(0)}$ is kept zero throughout the conduction zone.

$$\tilde{c}_{TZ}^{(0)} = \begin{cases} \alpha\left(1 - \dfrac{1}{\tilde{l}_D}\tilde{x}\right) & , \text{if } 0 \leq \tilde{x} \leq \tilde{l}_D, \\ 0 & , \text{if } \tilde{l}_D < \tilde{x} \leq 1, \end{cases} \qquad (52)$$

where $\tilde{l}_D$ is determined such that the concentration gradient drives $\tilde{j}^{(0)}$ in the diffusion zone by Equation (44): $\tilde{l}_D = \alpha/\tilde{j}^{(0)}$. $\alpha$ in the approximation can be obtained as a function of $\tilde{j}^{(0)}$ using the integral constraint in Equation (14). Since the triangular area under $\tilde{c}_{TZ}^{(0)}$ in the diffusion zone should be one, $\alpha = 2/\tilde{l}_D$, and therefore, $\alpha = \sqrt{2\tilde{j}^{(0)}}$ and $\tilde{l}_D = \sqrt{2/\tilde{j}^{(0)}}$. The approximated potential $\tilde{\phi}_{TZ}^{(0)}$ is then obtained by solving the conduction equation, $\tilde{\rho}\, d\tilde{\phi}_{TZ}^{(0)}/d\tilde{x} = \tilde{j}^{(0)}$, in the conduction zone. Since the potential drop in the diffusion zone is relatively small,

$$\tilde{\phi}_{TZ}^{(0)} = \begin{cases} 0 & , \text{if } 0 \leq \tilde{x} \leq \tilde{l}_D, \\ \dfrac{\tilde{j}^{(0)}}{\tilde{\rho}}\left(\tilde{x} - \tilde{l}_D\right) & , \text{if } \tilde{l}_D < \tilde{x} \leq 1. \end{cases} \qquad (53)$$

Though the expression is exactly the same to $\tilde{\phi}_{TZ}^{(0)}$ in the reservoir configuration, Equation (47), $\tilde{l}_D$ is a different function of $\tilde{j}^{(0)}$ between the configurations. $\tilde{c}_{TZ}^{(0)}$ and $\tilde{\phi}_{TZ}^{(0)}$ of the symmetric configuration are compared to the exact solutions in FIG 8 (d) and (e), respectively, where they show good agreement when $|\tilde{\rho}|$ is low and $\tilde{j}^{(0)}$ is large enough above the diffusion-limit.

The impedance approximation for the symmetric configuration is obtained also by the similar approach used in the reservoir configuration. First, $\tilde{c}_{TZ}^{(1)}$ is obtained by solving $d^2\tilde{c}_{TZ}^{(1)}/d\tilde{x}^2 = i\tilde{\omega}\tilde{c}_{TZ}^{(1)}$ in the diffusion zone with a boundary condition $d\tilde{c}_{TZ}^{(1)}/d\tilde{x} = -\tilde{j}^{(1)}$ at $\tilde{x} = \tilde{l}_D$ and an integral constraint $\int_0^{\tilde{l}_D} \tilde{c}_{TZ}^{(1)} d\tilde{x} = 0$.

$$\tilde{c}_{TZ}^{(1)} = -\tilde{j}^{(1)}\tilde{l}_D\left(\frac{\sinh\left(\sqrt{i\tilde{\omega}}\tilde{x}\right)}{\sqrt{i\tilde{\omega}}\tilde{l}_D} + \frac{\left(1 - \cosh\left(\sqrt{i\tilde{\omega}}\tilde{l}_D\right)\right)}{\sinh\left(\sqrt{i\tilde{\omega}}\tilde{l}_D\right)}\frac{\cosh\left(\sqrt{i\tilde{\omega}}\tilde{x}\right)}{\sqrt{i\tilde{\omega}}\tilde{l}_D}\right), \qquad (54)$$

for $0 \leq \tilde{x} \leq \tilde{l}_D$ (i.e., in the diffusion zone). Then, as we did for the reservoir configuration, $\tilde{\phi}^{(1)}/\tilde{c}^{(1)}$ at $\tilde{x} = \tilde{l}_D$ is calculated by solving the $O(\epsilon)$ terms of Equations (15) and (44), which turns out the same as that in the reservoir configuration shown in Equation (49). Following the steps described in Equations (50) and (51), the impedance approximation for the symmetric configuration is obtained:

$$\tilde{Z}_{TZ} = \tilde{Z}_{CZ} + \tilde{Z}_{DZ} = \frac{\tilde{l}_C}{-\tilde{\rho}} + \frac{\tilde{l}_D}{-\tilde{\rho}}\frac{\cosh\left(\sqrt{i\tilde{\omega}}\tilde{l}_D\right) - 1}{\sqrt{i\tilde{\omega}}\tilde{l}_D \sinh\left(\sqrt{i\tilde{\omega}}\tilde{l}_D\right)}, \qquad (55)$$

where $\tilde{l}_C = \left(\tilde{j}^{(0)} - \alpha\right)\big/\tilde{j}^{(0)}$, $\tilde{l}_D = \alpha\big/\tilde{j}^{(0)}$, and $\alpha = \sqrt{2\tilde{j}^{(0)}}$. Due to the integral constraint in Equation (14), $\tilde{Z}_{DZ}$ has a different expression in the symmetric configuration, even though it appears almost identical to the traditional finite-length Warburg element on the complex plane. It leads to the extra term in $\tilde{R}_L$ in Equation (41) compared to that of the reservoir configuration, which makes $\tilde{R}_L$ still increase in the overlimiting regime. In FIG 8 (f), $\tilde{Z}_{TZ}$ is compared to the exact $\tilde{Z}$ obtained numerically for the symmetric configuration. Similar to the reservoir configuration, the two-zone approximation provides an accurate approximation of $\tilde{Z}$ when $|\tilde{\rho}|$ is small and $\tilde{j}^{(0)}$ is large above the diffusion-limit.

Validity of the approximation largely depends on the interphase thickness ($\tilde{\delta}$) indicated in FIG 7 (c), because we assumed that the system can be exhaustively separated into the conduction and the diffusion zones by a sharp interphase between them. If $\tilde{\delta}$ is not small enough, contribution of the interphase should introduce a significant error in the approximation. We define $\tilde{\delta}$ by a distance between two positions where $\tilde{j}_D^{(0)} = 0.9\tilde{j}^{(0)}\big/(1-\tilde{\rho})$ and $\tilde{j}_C^{(0)} = 0.9\tilde{j}^{(0)}$. FIG 9 (a) and (b) show how $\tilde{\delta}$ changes by $\tilde{j}^{(0)}$ and $\tilde{\rho}$ in the overlimiting regime for the reservoir and the symmetric configurations, respectively. When $\tilde{j}^{(0)}$ is small and $|\tilde{\rho}|$ is large, there does not exist any distinct zones where either the conduction or the diffusion current dominates and the interphase thickness is not well defined. Otherwise, $\tilde{\delta}$ decreases in a direction of decreasing $|\tilde{\rho}|$ and increasing $\tilde{j}^{(0)}$. Its slope is steeper along $|\tilde{\rho}|$, and a sharp interface ($\tilde{\delta} < 0.1$) is obtained regardless of $\tilde{j}^{(0)}$, if $|\tilde{\rho}| < 0.01$ for the reservoir configuration or if $|\tilde{\rho}| < 0.03$ for the symmetric configuration, as long as an overlimiting bias is applied.

Error in the approximation is quantified by the average norm of relative residuals ($\Delta$) at a range of frequencies between $10^{-2}$ and $10^6$ spaced logarithmically with two points per decade.

$$\Delta = \frac{1}{N}\sum_{n=1}^{N}\left\|\frac{\tilde{Z}(\tilde{\omega}_n) - \tilde{Z}_{TZ}(\tilde{\omega}_n)}{|\tilde{Z}(\tilde{\omega}_n)|}\right\|, \quad (56)$$

where $N = 17$. FIG 9 (c) and (d) show $\Delta$ in a range of $\tilde{j}^{(0)}$ and $\tilde{\rho}$ for the reservoir and the symmetric configurations, respectively. $\Delta$ is relatively large above 0.1 in a region where $\tilde{j}^{(0)}$ is small and $|\tilde{\rho}|$ is large, when the two zones are not fully developed and $\tilde{\delta}$ is not defined. Then, it decreases quickly in a direction of decreasing $|\tilde{\rho}|$ and increasing $\tilde{j}^{(0)}$. A small $\Delta$ ($< 0.1$) is obtained even when the interface is somewhat thick with $\tilde{\delta}$ up to 0.3 at least, which is its largest value found in the examined range of $\tilde{j}^{(0)}$ and $\tilde{\rho}$. Therefore, it confirms that the two-zone approximation is valid as long as the two zones are developed and the interface is not too thick, which is achieved with a small $|\tilde{\rho}|$ and an overlimiting $\tilde{j}^{(0)}$.

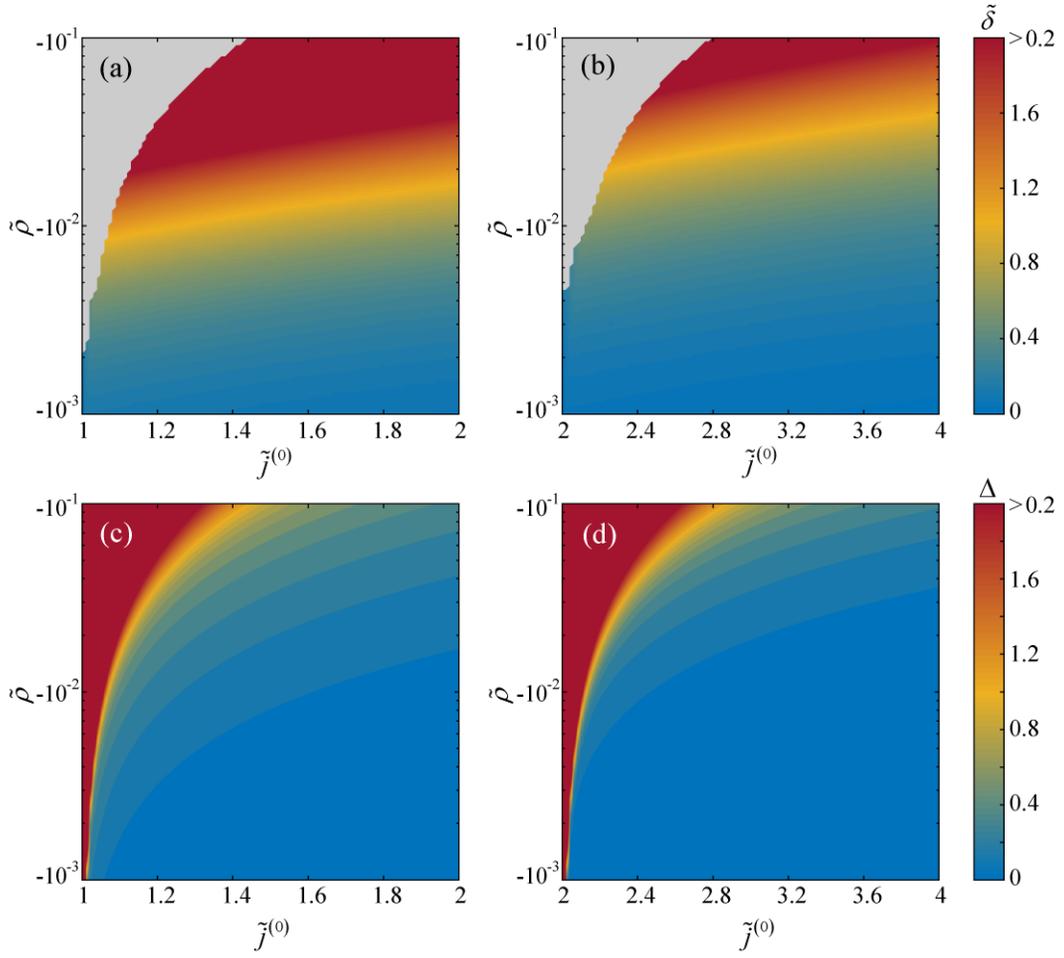

FIG 9. Validity of the two-zone approximation for a range of $\tilde{j}^{(0)}$ and $\tilde{\rho}$.
$\tilde{\delta}$ in (a) the reservoir and (b) the symmetric configurations.
$\Delta$ in (c) the reservoir and (d) the symmetric configurations.

## VI. CONCLUSION

The impedance of electrodiffusion in presence of a charged immobile species is studied by a theoretical model that combines the Nernst-Planck equations and the electroneutrality condition. In particular, its behavior is examined for a range of immobile charge density and applied bias, expanding to the overlimiting regime. The impedance of bulk electrodiffusion is commonly considered as a smaller contribution than those of interface kinetics. Under a *dc* bias, however, its magnitude increases exponentially, and it could be an important contribution to the overall cell impedance. In a biased condition, a bulk relaxation appears as a semicircle on the complex plane, which is attributed to perturbation in the bulk conductivity under the steady-state electric field.

As the applied *dc* bias approaches the diffusion-limit, the diffusion current dominates in the bulk and the impedance exhibits a transition to a series connection of a resistor and a finite-length Warburg element. When there is no charged immobile species or when they are co-charged to the active charge carriers, the impedance diverges before the diffusion-limiting bias, and an overlimiting bias may not be sustained by electrodiffusion unless promoted by other mechanisms. On the other hand, when the immobile species is counter-charged to the active charge carriers, an overlimiting bias can be applied. In the overlimiting regime with a further increasing bias, contribution of the resistor replaces that of the finite-length Warburg element, while the overall magnitude kept finite by the constant conductivity provided by charge carriers screening the immobile species.

Under an overlimiting bias, two distinct regions are found in the steady state solution with a sharp interphase. Based on the observation, we present an analytical approximation of the impedance where we assume complete dominance of conduction or diffusion in their respective zones. It attributes the resistor to the conduction zone and the finite-length Warburg element to the diffusion zone. The resistor is taking over the finite-length Warburg under increasing bias, because the conduction zone is replacing the diffusion zone. The approximation is valid when the immobile charge density is low and the bias is larger than the diffusion limit. Combined with an interface model [68], the present model and approximation can be employed to interpret impedance spectra of electrochemical cells with a charged immobile species under a *dc* bias even above the diffusion-limit.

**ACKNOWLEDGEMENT**

The authors acknowledge support by the Martin Family Fellowship and the Kwanjeong Fellowship to J. Song, the National Science Scholarship (PhD) funded by Agency for Science, Technology and Research, Singapore (A*STAR) to E. Khoo, and a Professor Amar G. Bose Research Grant to M. Z. Bazant.

# APPENDIX

## A. List of Symbols

| | |
|---|---|
| $c_+$ | Concentration of positive charge carrier |
| $c_-$ | Concentration of negative charge carrier |
| $c_0$ | Initial concentration of negative charge carrier |
| $\tilde{c}$ | Dimensionless concentration of charge carriers ($= c_-/c_0 = c_+/c_0 + 2\tilde{\rho}$) |
| $\tilde{c}^{(0)}$ | Steady state $\tilde{c}$ |
| $\tilde{c}^{(0)}_{TZ}$ | Two-zone approximation of $\tilde{c}^{(0)}$ |
| $\tilde{c}^{(1)}$ | Linear perturbation in $\tilde{c}$ |
| $\tilde{c}^{(1)}_{TZ}$ | Two-zone approximation of $\tilde{c}^{(1)}$ |
| $D$ | Diffusivity of charge carriers |
| $e$ | Electron charge number |
| $F_+$ | Flux of positive charge carrier |
| $F_-$ | Flux of negative charge carrier |
| $\tilde{F}_+$ | Dimensionless flux of positive charge carrier ($= LF_+/Dc_0$) |
| $\tilde{F}_-$ | Dimensionless flux of negative charge carrier ($= LF_-/Dc_0$) |
| $i$ | Unit imaginary number ($= \sqrt{-1}$) |
| $j$ | Current ($= ze(F_+ - F_-)$) |
| $j_{\lim}$ | Diffusion limiting current ($= 2zeDc_0/L$) |
| $\tilde{j}$ | Dimensionless current ($= j/j_{\lim}$) |
| $\tilde{j}^{(0)}$ | Steady state $dc$ bias in $\tilde{j}$ |
| $\tilde{j}^{(0)}_C$ | Conduction current in steady state ($= d\tilde{c}^{(0)}/d\tilde{x}$) |
| $\tilde{j}^{(0)}_D$ | Diffusion current in steady state ($= -\tilde{\rho}\, d\tilde{\phi}^{(0)}/d\tilde{x}$) |
| $\tilde{j}^{(1)}$ | Linear perturbation in $\tilde{j}$ |
| $k_B$ | Boltzmann constant |
| $\tilde{l}_D$ | Dimensionless length of diffusion zone |
| $\tilde{l}_C$ | Dimensionless length of conduction zone |
| $L$ | Length of the system |
| $\tilde{R}_H$ | High-frequency-limit resistance |
| $\tilde{R}_L$ | Low-frequency-limit resistance |
| $t$ | Time variable |
| $\tilde{t}$ | Dimensionless time variable ($= Dt/L^2$) |
| $T$ | Temperature |
| $V$ | Applied potential |

| | | |
|---|---|---|
| $\tilde{V}$ | Dimensionless applied potential ($= zeV/k_BT$) | |
| $\tilde{V}^{(0)}$ | dc bias in $\tilde{V}$ | |
| $\tilde{V}^{(1)}$ | Linear perturbation in $\tilde{V}$ | |
| $x$ | Position variable | |
| $\tilde{x}$ | Dimensionless position variable ($= x/L$) | |
| $z$ | Charge number of charge carriers | |
| $Z$ | Impedance | |
| $\tilde{Z}$ | Dimensionless impedance ($= 2Dc_0Z/k_BTL$) | |
| $\tilde{Z}_{CZ}$ | Dimensionless impedance of conduction zone | |
| $\tilde{Z}_{DZ}$ | Dimensionless impedance of diffusion zone | |
| $\tilde{Z}_{TZ}$ | Two-zone approximation of $\tilde{Z}$ | |

*Greeks*

| | | |
|---|---|---|
| $\alpha$ | Intermediate parameter, defined in Equation (23) | |
| $\tilde{\delta}$ | Interphase thickness, dimensionless | |
| $\Delta$ | Average norm of relative residuals | |
| $\epsilon$ | Arbitrary small number | |
| $\theta_H$ | High-frequency-limit local phase angle | |
| $\rho$ | Charge density of immobile species | |
| $\tilde{\rho}$ | Dimensionless charge density of immobile species ($= \rho/2zec_0$) | |
| $\phi$ | Electric potential | |
| $\tilde{\phi}$ | Dimensionless electric potential ($= ze\phi/k_BT$) | |
| $\tilde{\phi}^{(0)}$ | Steady state $\tilde{\phi}$ | |
| $\tilde{\phi}^{(0)}_{TZ}$ | Two-zone approximation of $\tilde{\phi}^{(0)}$ | |
| $\tilde{\phi}^{(1)}$ | Linear perturbation in $\tilde{\phi}$ | |
| $\tilde{\phi}^{(1)}_{TZ}$ | Two-zone approximation of $\tilde{\phi}^{(1)}$ | |
| $\omega$ | Frequency of perturbation | |
| $\tilde{\omega}$ | Dimensionless frequency of perturbation ($= \omega L^2/D$) | |

## B. Analytical Solutions in Steady State

The steady state solution can be obtained by solving the $O(1)$ terms in the perturbation. Collecting the $O(1)$ terms, the governing equations become

$$0 = \frac{d^2 \tilde{c}^{(0)}}{d\tilde{x}^2} - \tilde{\rho} \frac{d^2 \tilde{\phi}^{(0)}}{d\tilde{x}^2}, \tag{A1}$$

$$0 = \frac{d}{d\tilde{x}} \left( \left( \tilde{c}^{(0)} - \tilde{\rho} \right) \frac{d\tilde{\phi}^{(0)}}{d\tilde{x}} \right). \tag{A2}$$

Equations (A1) and (A2) can be integrated once by introducing the current ($\tilde{j}^{(0)}$), which is unknown at this stage but will be determined later for each configuration. Since the system is locally neutral, $\tilde{j}^{(0)}$ is constant throughout the domain. The current is entirely carried by the positive charge carrier everywhere in steady state, since the negative charge carrier is blocked at the boundary, $\tilde{x} = 1$. Then, the total flux of charge carriers as well as the net charge current should be $\tilde{j}^{(0)}$ in $O(1)$ dimensionless form. Thereby

$$-\tilde{j}^{(0)} = \frac{d\tilde{c}^{(0)}}{d\tilde{x}} - \tilde{\rho} \frac{d\tilde{\phi}^{(0)}}{d\tilde{x}}, \tag{A3}$$

$$-\tilde{j}^{(0)} = \left( \tilde{c}^{(0)} - \tilde{\rho} \right) \frac{d\tilde{\phi}^{(0)}}{d\tilde{x}}. \tag{A4}$$

By subtracting Equation (A4) from Equation (A3), we recover the flux of the negative charge carriers, which should be zero.

$$0 = \frac{d\tilde{c}^{(0)}}{d\tilde{x}} - \tilde{c}^{(0)} \frac{d\tilde{\phi}^{(0)}}{d\tilde{x}}. \tag{A5}$$

Equations (A3) and (A5) are more straightforward to solve analytically.

For the reservoir configuration, we first integrate Equation (A5), and apply the boundary condition at $\tilde{x} = 0$: $\tilde{c}^{(0)} = 1$ and $\tilde{\phi}^{(0)} = 0$. Then, we obtain

$$\log\left( \tilde{c}^{(0)} \right) = \tilde{\phi}^{(0)}. \tag{A6}$$

Equation (A3) can be integrated using the boundary condition at $\tilde{x} = 0$ again.

$$1 - \tilde{j}^{(0)} \tilde{x} = \tilde{c}^{(0)} - \tilde{\rho} \tilde{\phi}^{(0)}. \tag{A7}$$

Then, $\tilde{j}^{(0)}$ can be determined by the boundary condition at $\tilde{x} = 1$: $\tilde{\phi}^{(0)} = -\tilde{V}^{(0)}$.

$$\tilde{j}^{(0)} = 1 - \left(e^{-\tilde{V}^{(0)}} + \tilde{\rho}\tilde{V}^{(0)}\right). \tag{A8}$$

Equations (A6), (A7), and (A8) make up an implicit form of the steady state solution for the reservoir configuration. They become explicit if $\tilde{\rho} = 0$. To obtain an explicit form for any values of $\tilde{\rho}$, we plug-in Equation (A6) in to Equation (A7) and rearrange to

$$\frac{1}{-\tilde{\rho}} \exp\left(\frac{1 - \tilde{j}^{(0)}\tilde{x}}{-\tilde{\rho}}\right) = \frac{\tilde{c}^{(0)}}{-\tilde{\rho}} \exp\left(\frac{\tilde{c}^{(0)}}{-\tilde{\rho}}\right). \tag{A9}$$

Then, we recognize that $\left(-\tilde{c}^{(0)}/\tilde{\rho}\right)$ should be the Lambert W function, unless $\tilde{\rho} = 0$. Depending on sign of $\tilde{\rho}$, a different branch gives a physically meaningful solution. When $\tilde{\rho} = 0$, Equation (A7) gives the explicit solution for $\tilde{c}^{(0)}$. Collectively, we obtain Equation (18) as for an explicit form of the steady state solution for the reservoir configuration for any values of $\tilde{\rho}$.

On the other hand for the symmetric configuration, we again integrate Equation (A5). We introduce an unknown parameter $\alpha = \tilde{c}^{(0)}(\tilde{x} = 0)$, which will be determined later. By applying the boundary conditions at $\tilde{x} = 0$, $\tilde{c}^{(0)} = \alpha$ and $\tilde{\phi}^{(0)} = 0$, we obtain

$$\log\left(\frac{\tilde{c}^{(0)}}{\alpha}\right) = \tilde{\phi}^{(0)}. \tag{A10}$$

Equation (A3) can be integrated using the boundary conditions at $\tilde{x} = 0$ again.

$$\alpha - \tilde{j}^{(0)}\tilde{x} = \tilde{c}^{(0)} - \tilde{\rho}\tilde{\phi}^{(0)}. \tag{A11}$$

Then $\tilde{j}^{(0)}$ can be determined by the boundary condition at $\tilde{x} = 1$, $\tilde{\phi}^{(0)} = -\tilde{V}^{(0)}$.

$$\tilde{j}^{(0)} = \alpha\left(1 - e^{-\tilde{V}^{(0)}}\right) - \tilde{\rho}\tilde{V}^{(0)}. \tag{A12}$$

Upon determining $\alpha$, Equations (A10), (A11), and (A12) provide an implicit form of the steady state solution for the symmetric configuration. They become explicit if $\tilde{\rho} = 0$. $\alpha$ can be determined by employing the integral constraint:

$$\int_0^1 \tilde{c}^{(0)} d\tilde{x} = \alpha \int_0^{-\tilde{V}^{(0)}} e^{\tilde{\phi}^{(0)}} \left(\frac{d\tilde{\phi}^{(0)}}{d\tilde{x}}\right)^{-1} d\tilde{\phi}^{(0)} = 1, \tag{A13}$$

where $d\tilde{\phi}^{(0)}/d\tilde{x}$ can be obtained by combining (A12) and (A3).

$$-\alpha e^{\tilde{\phi}^{(0)}}\frac{d\tilde{\phi}^{(0)}}{d\tilde{x}} + \tilde{\rho}\frac{d\tilde{\phi}^{(0)}}{d\tilde{x}} = \alpha\left(1 - e^{-\tilde{V}^{(0)}}\right) - \tilde{\rho}\tilde{V}^{(0)}, \tag{A14}$$

$$\frac{d\tilde{\phi}^{(0)}}{d\tilde{x}} = \frac{\alpha\left(1-e^{-\tilde{V}^{(0)}}\right)-\tilde{\rho}\tilde{V}^{(0)}}{\tilde{\rho}-\alpha e^{\tilde{\phi}^{(0)}}}. \tag{A15}$$

By plugging-in Equation (A15) to Equation (A13), and rearranging the terms,

$$\alpha\int_0^{-\tilde{V}^{(0)}} \left(\tilde{\rho}e^{\tilde{\phi}^{(0)}} - \alpha e^{2\tilde{\phi}^{(0)}}\right)d\tilde{\phi}^{(0)} = \alpha\left(1-e^{-\tilde{V}^{(0)}}\right)-\tilde{\rho}\tilde{V}^{(0)}. \tag{A16}$$

Upon performing the integrations, Equation (A16) becomes a quadratic equation for $\alpha$:

$$\left(1-e^{-2\tilde{V}^{(0)}}\right)\alpha^2 + 2(1+\tilde{\rho})\left(e^{-\tilde{V}^{(0)}}-1\right)\alpha + 2\tilde{\rho}\tilde{V}^{(0)} = 0. \tag{A17}$$

Only one of the roots provides a physically-meaningful solution. Therefore,

$$\alpha = \frac{(1+\tilde{\rho})\left(1-e^{-\tilde{V}^{(0)}}\right)+\sqrt{(1+\tilde{\rho})^2\left(1-e^{-\tilde{V}^{(0)}}\right)^2-2\tilde{\rho}\tilde{V}^{(0)}\left(1-e^{-2\tilde{V}^{(0)}}\right)}}{1-e^{-2\tilde{V}^{(0)}}}. \tag{A18}$$

To obtain an explicit form for any values of $\tilde{\rho}$, we plug-in Equation (A10) in to Equation (A11) and rearrange to

$$\frac{\alpha}{-\tilde{\rho}}\exp\left(\frac{\alpha-\tilde{j}^{(0)}\tilde{x}}{-\tilde{\rho}}\right) = \frac{\tilde{c}^{(0)}}{-\tilde{\rho}}\exp\left(\frac{\tilde{c}^{(0)}}{-\tilde{\rho}}\right). \tag{A19}$$

Same to the reservoir configuration, we recognize that $\left(-\tilde{c}^{(0)}/\tilde{\rho}\right)$ should be the Lambert W function, unless $\tilde{\rho}=0$. Depending on sign of $\tilde{\rho}$, a different branch gives a physically meaningful solution. When $\tilde{\rho}=0$, Equation (A11) gives the explicit solution for $\tilde{c}^{(0)}$. Collectively, we obtain Equation (21) as for an explicit form of the steady state solution for the symmetric configuration for any values of $\tilde{\rho}$.

## C. Small-bias Limits of $\tilde{R}_L$ and $\tilde{R}_H$

Under a small bias, $\tilde{c}^{(0)}(\tilde{x})$ can be approximated by a linear function. To satisfy the boundary conditions for the reservoir configuration,

$$\tilde{c}^{(0)}\left(\tilde{j}^{(0)}\ll 1\right) \simeq 1 - \frac{\tilde{j}^{(0)}}{1-\tilde{\rho}}\tilde{x}. \tag{B1}$$

In the low-frequency limit, the impedance should converge to the local resistance in steady state, i.e., $\tilde{R}_L = d\tilde{V}^{(0)}/d\tilde{j}^{(0)}$. Under a small bias, $\tilde{\phi}^{(0)}(\tilde{x})$ is obtained by integrating the $O(1)$ terms of Equation (15): $-\tilde{j}^{(0)} = \left(\tilde{c}^{(0)}-\tilde{\rho}\right)d\tilde{\phi}^{(0)}/d\tilde{x}$, using $\tilde{c}^{(0)}$ in Equation (B1) and a boundary condition, $\tilde{\phi}^{(0)}=0$ at $\tilde{x}=0$.

$$\tilde{\phi}^{(0)}\left(\tilde{j}^{(0)} \ll 1\right) \simeq (1-\tilde{\rho})\log\left(\frac{(1-\tilde{\rho})^2 - \tilde{j}^{(0)}\tilde{x}}{(1-\tilde{\rho})^2}\right). \tag{B2}$$

Then $\tilde{V}^{(0)}$ is obtained by $-\tilde{\phi}^{(0)}$ at $\tilde{x}=1$.

$$\tilde{V}^{(0)}\left(\tilde{j}^{(0)} \ll 1\right) \simeq -(1-\tilde{\rho})\log\left(\frac{(1-\tilde{\rho})^2 - \tilde{j}^{(0)}}{(1-\tilde{\rho})^2}\right). \tag{B3}$$

Therefore,

$$\tilde{R}_L\left(\tilde{j}^{(0)} \ll 1\right) \simeq \frac{(1-\tilde{\rho})}{(1-\tilde{\rho})^2 - \tilde{j}^{(0)}}. \tag{B4}$$

In the high-frequency limit, $\tilde{c}^{(1)}$ does not respond to the perturbation, and $\tilde{\phi}^{(1)}$ is obtained by solving the $O(\epsilon)$ terms of Equation (15) with $\tilde{c}^{(1)} = 0$: $-\tilde{j}^{(1)} = \left(\tilde{c}^{(0)} - \tilde{\rho}\right)d\tilde{\phi}^{(1)}/d\tilde{x}$. Then evaluating $-\tilde{\phi}^{(1)}$ at $\tilde{x}=1$, $\tilde{V}^{(1)}$ becomes

$$\tilde{V}^{(1)}\left(\tilde{j}^{(0)} \ll 1\right) \simeq -\frac{\tilde{j}^{(1)}(1-\tilde{\rho})}{\tilde{j}^{(0)}}\log\left(\frac{(1-\tilde{\rho})^2 - \tilde{j}^{(0)}}{(1-\tilde{\rho})^2}\right). \tag{B5}$$

Then, $\tilde{R}_H$ under a small bias is obtained by

$$\tilde{R}_H\left(\tilde{j}^{(0)} \ll 1\right) = \frac{\tilde{V}^{(1)}\left(\tilde{j}^{(0)} \ll 1\right)}{\tilde{j}^{(1)}} \simeq -\frac{(1-\tilde{\rho})}{\tilde{j}^{(0)}}\log\left(\frac{(1-\tilde{\rho})^2 - \tilde{j}^{(0)}}{(1-\tilde{\rho})^2}\right). \tag{B6}$$

The same approach can be applied to the symmetric configuration under a small bias ($\tilde{j}^{(0)} \ll 2$). First, $\tilde{c}^{(0)}(\tilde{x})$ is approximated by a linear function.

$$\tilde{c}^{(0)}\left(\tilde{j}^{(0)} \ll 2\right) \simeq \alpha - \frac{\tilde{j}^{(0)}}{1-\tilde{\rho}}\tilde{x}, \tag{B7}$$

where $\alpha$ can be obtained as a function of $\tilde{j}^{(0)}$ employing the integral constraint in Equation (14): $\alpha\left(\tilde{j}^{(0)} \ll 2\right) \simeq 1 + \tilde{j}/(2-2\tilde{\rho})$. Then $\tilde{\phi}^{(0)}(\tilde{x})$ and $\tilde{V}^{(0)} = -\tilde{\phi}^{(0)}(1)$ are obtained:

$$\tilde{\phi}^{(0)}\left(\tilde{j}^{(0)} \ll 2\right) \simeq (1-\tilde{\rho})\log\left(\frac{(\alpha-\tilde{\rho})(1-\tilde{\rho}) - \tilde{j}^{(0)}\tilde{x}}{(\alpha-\tilde{\rho})(1-\tilde{\rho})}\right), \tag{B8}$$

$$\tilde{V}^{(0)}\left(\tilde{j}^{(0)} \ll 2\right) \simeq -(1-\tilde{\rho})\log\left(\frac{-\tilde{j}^{(0)} + 2(1-\tilde{\rho})^2}{\tilde{j}^{(0)} + 2(1-\tilde{\rho})^2}\right). \tag{B9}$$

$\tilde{R}_L$ is obtained by $d\tilde{V}^{(0)}/d\tilde{j}^{(0)}$. For the symmetric condition under a small bias,

$$\tilde{R}_L\left(\tilde{j}^{(0)} \ll 2\right) \simeq \frac{(1-\tilde{\rho})}{2(1-\tilde{\rho})^2 - \tilde{j}^{(0)}} + \frac{(1-\tilde{\rho})}{2(1-\tilde{\rho})^2 + \tilde{j}^{(0)}}. \tag{B10}$$

Then to obtain $\tilde{R}_H$, $\tilde{V}^{(1)}\left(\tilde{j}^{(0)} \ll 2\right)$ is obtained by solving $-\tilde{j}^{(1)} = \left(\tilde{c}^{(0)} - \tilde{\rho}\right) d\tilde{\phi}^{(1)}/d\tilde{x}$ for $\tilde{\phi}^{(1)}$, and evaluating it at $\tilde{x}=1$.

$$\tilde{V}^{(1)}\left(\tilde{j}^{(0)} \ll 2\right) \simeq -\frac{\tilde{j}^{(1)}(1-\tilde{\rho})}{\tilde{j}^{(0)}} \log\left(\frac{-\tilde{j}^{(0)} + 2(1-\tilde{\rho})^2}{\tilde{j}^{(0)} + 2(1-\tilde{\rho})^2}\right). \tag{B11}$$

$\tilde{R}_H$ is obtained by $\tilde{V}^{(1)}/\tilde{j}^{(1)}$.

$$\tilde{R}_H\left(\tilde{j}^{(0)} \ll 2\right) \simeq -\frac{(1-\tilde{\rho})}{\tilde{j}^{(0)}} \log\left(\frac{-\tilde{j}^{(0)} + 2(1-\tilde{\rho})^2}{\tilde{j}^{(0)} + 2(1-\tilde{\rho})^2}\right). \tag{B12}$$

The small-bias asymptotic expressions of $\tilde{R}_L$ and $\tilde{R}_H$ are compared to the numerical limits of the full model in FIG A1. For both configurations, the asymptotic expressions show good agreement in the small-bias regime, even close to the limiting current.

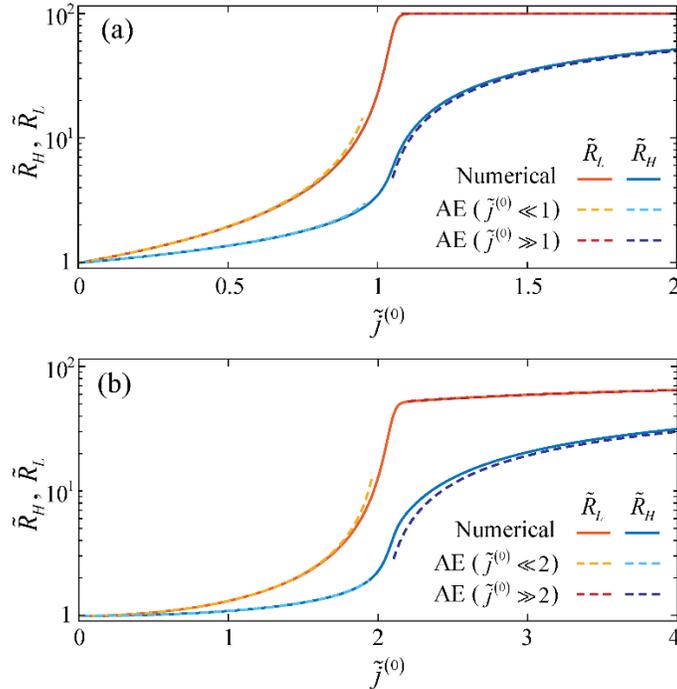

FIG A1. Comparison of the asymptotic expressions (AEs) of $\tilde{R}_L$ and $\tilde{R}_H$ to the numerical limits of the full model: (a) the reservoir configuration and (b) the symmetric configuration.

## D. Large-bias Limits of $\tilde{R}_L$ and $\tilde{R}_H$

Asymptotic expressions under a large bias above the diffusion limit can be obtained from the two-zone approximation in Section V. $\tilde{R}_L$ is again obtained by the local resistance in the steady-state approximation. For the reservoir configuration, $\tilde{\phi}_{TZ}^{(0)}$ in Equation (47) is evaluated at $\tilde{x}=1$.

$$\tilde{V}^{(0)}\left(\tilde{j}^{(0)} \gg 1\right) \simeq -\frac{\tilde{j}^{(0)}}{\tilde{\rho}}\left(1-\tilde{l}_D\right), \tag{C1}$$

where $\tilde{l}_D = 1/\tilde{j}^{(0)}$. Therefore, $d\tilde{V}^{(0)}/d\tilde{j}^{(0)}$ leads to

$$\tilde{R}_L\left(\tilde{j}^{(0)} \gg 1\right) \simeq \frac{1}{-\tilde{\rho}}. \tag{C2}$$

In the high-frequency limit, the diffusion zone contributes little to the impedance. Therefore, $\tilde{R}_H$ comes only from the impedance of the conduction zone, $\tilde{Z}_{CZ}$. Therefore,

$$\tilde{R}_H \simeq \frac{\tilde{l}_C}{-\tilde{\rho}} = \frac{1}{-\tilde{\rho}} \frac{\left(\tilde{j}^{(0)}-1\right)}{\tilde{j}^{(0)}}. \tag{C3}$$

The same approach can be applied to the symmetric configuration under a large bias ($\tilde{j}^{(0)} \gg 2$). Evaluating $\tilde{\phi}_{TZ}^{(0)}$ in Equation (53) at $\tilde{x}=1$,

$$\tilde{V}^{(0)}\left(\tilde{j}^{(0)} \gg 2\right) \simeq -\frac{\tilde{j}^{(0)}}{\tilde{\rho}}\left(1-\tilde{l}_D\right), \tag{C4}$$

where $\tilde{l}_D = \sqrt{2/\tilde{j}^{(0)}}$ in the approximation. Then, $d\tilde{V}^{(0)}/d\tilde{j}^{(0)}$ leads to

$$\tilde{R}_L\left(\tilde{j}^{(0)} \gg 2\right) \simeq \frac{1}{-\tilde{\rho}}\left(1-\frac{1}{\sqrt{2\tilde{j}^{(0)}}}\right). \tag{C5}$$

$\tilde{R}_H$ comes only from $\tilde{Z}_{CZ}$. For the symmetric configuration,

$$\tilde{R}_H\left(\tilde{j}^{(0)} \gg 2\right) \simeq \frac{1}{-\tilde{\rho}} \frac{\tilde{j}^{(0)}-\sqrt{2\tilde{j}^{(0)}}}{\tilde{j}^{(0)}}. \tag{C6}$$

The large-bias asymptotic expressions of $\tilde{R}_L$ and $\tilde{R}_H$ are compared to the numerical limits of the full model in FIG A1. Like the small-bias expressions, the large-bias expressions show good agreement in the overlimiting regime, even close to the limiting current when $\tilde{\rho}=-0.01$. Since they are based on the two-zone approximation (Section V), they are valid when $|\tilde{\rho}|$ is low and the bias is above the diffusion limit.